\documentclass[12pt,a4paper]{article}
\pdfoutput=1

\usepackage{ifthen} 
\newboolean{pdflatex}
\setboolean{pdflatex}{true} 

\newboolean{articletitles}
\setboolean{articletitles}{true} 

\newboolean{uprightparticles}
\setboolean{uprightparticles}{false} 

\newboolean{inbibliography}
\setboolean{inbibliography}{false} 

\usepackage{microtype}
\usepackage{lineno}  
\usepackage{xspace} 
\usepackage{caption} 

\usepackage{graphicx}  
\usepackage{color}
\usepackage{colortbl}
\graphicspath{{./figs/}{../../ANANOTE/latex/figs/}} 

\usepackage{amsmath} 
\usepackage{amssymb}
\usepackage{amsfonts}
\usepackage{upgreek} 

\newcommand*\patchAmsMathEnvironmentForLineno[1]{%
\expandafter\let\csname old#1\expandafter\endcsname\csname #1\endcsname
\expandafter\let\csname oldend#1\expandafter\endcsname\csname
end#1\endcsname
 \renewenvironment{#1}%
   {\linenomath\csname old#1\endcsname}%
   {\csname oldend#1\endcsname\endlinenomath}%
}
\newcommand*\patchBothAmsMathEnvironmentsForLineno[1]{%
  \patchAmsMathEnvironmentForLineno{#1}%
  \patchAmsMathEnvironmentForLineno{#1*}%
}
\AtBeginDocument{%
\patchBothAmsMathEnvironmentsForLineno{equation}%
\patchBothAmsMathEnvironmentsForLineno{align}%
\patchBothAmsMathEnvironmentsForLineno{flalign}%
\patchBothAmsMathEnvironmentsForLineno{alignat}%
\patchBothAmsMathEnvironmentsForLineno{gather}%
\patchBothAmsMathEnvironmentsForLineno{multline}%
\patchBothAmsMathEnvironmentsForLineno{eqnarray}%
}

\usepackage{hyperref}    
\usepackage[all]{hypcap} 

\usepackage{xspace} 
\usepackage{upgreek}

\def\lhcb {\mbox{LHCb}\xspace}

\def\babar  {\mbox{BaBar}\xspace}
\def\belle  {\mbox{Belle}\xspace}

\def\MagUp {\mbox{\em Mag\kern -0.05em Up}\xspace}

\ifthenelse{\boolean{uprightparticles}}%
{

 \def\Ppi         {\ensuremath{\uppi}\xspace}

 \def\Ppsi        {\ensuremath{\uppsi}\xspace}

 \def\PDelta      {\ensuremath{\Delta}\xspace}                 
 \def\PXi      {\ensuremath{\Xi}\xspace}                 
 \def\PLambda      {\ensuremath{\Lambda}\xspace}                 
 \def\PSigma      {\ensuremath{\Sigma}\xspace}                 
 \def\POmega      {\ensuremath{\Omega}\xspace}                 
 \def\PUpsilon      {\ensuremath{\Upsilon}\xspace}                 
 
 \def\PB      {\ensuremath{\mathrm{B}}\xspace}                 
                  
 \def\PD      {\ensuremath{\mathrm{D}}\xspace}

 \def\PJ      {\ensuremath{\mathrm{J}}\xspace}                 
 \def\PK      {\ensuremath{\mathrm{K}}\xspace}

 \def\Pb      {\ensuremath{\mathrm{b}}\xspace}                 
 \def\Pc      {\ensuremath{\mathrm{c}}\xspace}                 
 \def\Pd      {\ensuremath{\mathrm{d}}\xspace}

 \def\Pi      {\ensuremath{\mathrm{i}}\xspace}

 \def\Pp      {\ensuremath{\mathrm{p}}\xspace}

 \def\Ps      {\ensuremath{\mathrm{s}}\xspace}

}
{

 \def\Ppi         {\ensuremath{\pi}\xspace}

 \def\Ppsi        {\ensuremath{\psi}\xspace}                 
                  
 \mathchardef\PDelta="7101
 \mathchardef\PXi="7104
 \mathchardef\PLambda="7103
 \mathchardef\PSigma="7106
 \mathchardef\POmega="710A
 \mathchardef\PUpsilon="7107
                  
 \def\PB      {\ensuremath{B}\xspace}                 
                  
 \def\PD      {\ensuremath{D}\xspace}

 \def\PJ      {\ensuremath{J}\xspace}                 
 \def\PK      {\ensuremath{K}\xspace}

 \def\Pb      {\ensuremath{b}\xspace}                 
 \def\Pc      {\ensuremath{c}\xspace}                 
 \def\Pd      {\ensuremath{d}\xspace}

 \def\Pi      {\ensuremath{i}\xspace}

 \def\Pp      {\ensuremath{p}\xspace}

 \def\Ps      {\ensuremath{s}\xspace}

}

\makeatletter
\ifcase \@ptsize \relax
  \newcommand{\miniscule}{\@setfontsize\miniscule{4}{5}}
\or
  \newcommand{\miniscule}{\@setfontsize\miniscule{5}{6}}
\or
  \newcommand{\miniscule}{\@setfontsize\miniscule{5}{6}}
\fi
\makeatother

\DeclareRobustCommand{\optbar}[1]{\shortstack{{\miniscule (\rule[.5ex]{1.25em}{.18mm})}
  \\ [-.7ex] $#1$}}

\def\dquark    {{\ensuremath{\Pd}}\xspace}

\def\squark    {{\ensuremath{\Ps}}\xspace}
\def\squarkbar {{\ensuremath{\overline \squark}}\xspace}
\def\ssbar     {{\ensuremath{\squark\squarkbar}}\xspace}
\def\cquark    {{\ensuremath{\Pc}}\xspace}

\def\bquark    {{\ensuremath{\Pb}}\xspace}

\def\pion   {{\ensuremath{\Ppi}}\xspace}

\def\pip    {{\ensuremath{\pion^+}}\xspace}
\def\pim    {{\ensuremath{\pion^-}}\xspace}

\def\kaon    {{\ensuremath{\PK}}\xspace}
  \def\Kbar    {{\kern 0.2em\overline{\kern -0.2em \PK}{}}\xspace}

\def\KorKbar    {\kern 0.18em\optbar{\kern -0.18em K}{}\xspace}
\def\Kz      {{\ensuremath{\kaon^0}}\xspace}

\def\Kp      {{\ensuremath{\kaon^+}}\xspace}
\def\Km      {{\ensuremath{\kaon^-}}\xspace}

\def\KS      {{\ensuremath{\kaon^0_{\mathrm{ \scriptscriptstyle S}}}}\xspace}

\def\Kstarz  {{\ensuremath{\kaon^{*0}}}\xspace}

  \def\Dbar    {{\kern 0.2em\overline{\kern -0.2em \PD}{}}\xspace}
\def\D       {{\ensuremath{\PD}}\xspace}

\def\DorDbar    {\kern 0.18em\optbar{\kern -0.18em D}{}\xspace}
\def\Dz      {{\ensuremath{\D^0}}\xspace}

\def\B       {{\ensuremath{\PB}}\xspace}
\def\Bbar    {{\ensuremath{\kern 0.18em\overline{\kern -0.18em \PB}{}}}\xspace}

\def\BorBbar    {\kern 0.18em\optbar{\kern -0.18em B}{}\xspace}

\def\Bu      {{\ensuremath{\B^+}}\xspace}

\def\Bd      {{\ensuremath{\B^0}}\xspace}
\def\Bs      {{\ensuremath{\B^0_\squark}}\xspace}

\def\jpsi     {{\ensuremath{{\PJ\mskip -3mu/\mskip -2mu\Ppsi\mskip 2mu}}}\xspace}

  \def\Y#1S{\ensuremath{\PUpsilon{(#1S)}}\xspace}

\def\proton      {{\ensuremath{\Pp}}\xspace}

\def\Lz          {{\ensuremath{\PLambda}}\xspace}
\def\Lbar        {{\ensuremath{\kern 0.1em\overline{\kern -0.1em\PLambda}}}\xspace}
\def\LorLbar    {\kern 0.18em\optbar{\kern -0.18em \PLambda}{}\xspace}

\def\Lb      {{\ensuremath{\Lz^0_\bquark}}\xspace}

\def\BF         {{\ensuremath{\mathcal{B}}}\xspace}

\def\BR         {\BF}
\newcommand{\decay}[2]{\ensuremath{#1\!\to #2}\xspace}         

\def\to                 {\ensuremath{\rightarrow}\xspace}

\def\CP                {{\ensuremath{C\!P}}\xspace}

\def\AT#1     {\ensuremath{A_{\mathrm{T}}^{#1}}\xspace}           

\def\C#1      {\ensuremath{\mathcal{C}_{#1}}\xspace}                       
\def\Cp#1     {\ensuremath{\mathcal{C}_{#1}^{'}}\xspace}                    
\def\Ceff#1   {\ensuremath{\mathcal{C}_{#1}^{\mathrm{(eff)}}}\xspace}        
\def\Cpeff#1  {\ensuremath{\mathcal{C}_{#1}^{'\mathrm{(eff)}}}\xspace}       
\def\Ope#1    {\ensuremath{\mathcal{O}_{#1}}\xspace}                       
\def\Opep#1   {\ensuremath{\mathcal{O}_{#1}^{'}}\xspace}                    

\newcommand{\tev}{\ifthenelse{\boolean{inbibliography}}{\ensuremath{~T\kern -0.05em eV}\xspace}{\ensuremath{\mathrm{\,Te\kern -0.1em V}}}\xspace}
\newcommand{\gev}{\ensuremath{\mathrm{\,Ge\kern -0.1em V}}\xspace}
\newcommand{\mev}{\ensuremath{\mathrm{\,Me\kern -0.1em V}}\xspace}
\newcommand{\kev}{\ensuremath{\mathrm{\,ke\kern -0.1em V}}\xspace}
\newcommand{\ev}{\ensuremath{\mathrm{\,e\kern -0.1em V}}\xspace}
\newcommand{\gevc}{\ensuremath{{\mathrm{\,Ge\kern -0.1em V\!/}c}}\xspace}
\newcommand{\mevc}{\ensuremath{{\mathrm{\,Me\kern -0.1em V\!/}c}}\xspace}
\newcommand{\gevcc}{\ensuremath{{\mathrm{\,Ge\kern -0.1em V\!/}c^2}}\xspace}
\newcommand{\gevgevcccc}{\ensuremath{{\mathrm{\,Ge\kern -0.1em V^2\!/}c^4}}\xspace}
\newcommand{\mevcc}{\ensuremath{{\mathrm{\,Me\kern -0.1em V\!/}c^2}}\xspace}

\def\mum  {\ensuremath{{\,\upmu\mathrm{m}}}\xspace}

\def\invfb   {\ensuremath{\mbox{\,fb}^{-1}}\xspace}

\newcommand{\stat}{\ensuremath{\mathrm{\,(stat)}}\xspace}
\newcommand{\syst}{\ensuremath{\mathrm{\,(syst)}}\xspace}

\newcommand{\chisq}{\ensuremath{\chi^2}\xspace}

\def\gsim{{~\raise.15em\hbox{$>$}\kern-.85em
          \lower.35em\hbox{$\sim$}~}\xspace}
\def\lsim{{~\raise.15em\hbox{$<$}\kern-.85em
          \lower.35em\hbox{$\sim$}~}\xspace}

\def\sPlot{\mbox{\em sPlot}\xspace}

\def\ptot       {\mbox{$p$}\xspace}
\def\pt         {\mbox{$p_{\mathrm{ T}}$}\xspace}

\def\mrad{\ensuremath{\mathrm{ \,mrad}}\xspace}

\def\evtgen     {\mbox{\textsc{EvtGen}}\xspace}

\def\geant      {\mbox{\textsc{Geant4}}\xspace}

\def\photos     {\mbox{\textsc{Photos}}\xspace}

\def\pythiaeight     {\mbox{\textsc{Pythia8}}\xspace}

\def\tell1  {TELL1\xspace}
\def\ukl1   {UKL1\xspace}

\newcommand{\ie}{\mbox{\itshape i.e.}\xspace}

\newcommand{\LbF}{\ensuremath{{\proton\pim\Kp\Km}\xspace}}
\newcommand{\LT}{\ensuremath{{\proton\pim}\xspace}}
\newcommand{\BdF}{\ensuremath{{\pip\pim\Kp\Km}\xspace}}
\newcommand{\KST}{\ensuremath{{\pip\pim}\xspace}}
\newcommand{\phiT}{\ensuremath{{\Kp\Km}\xspace}}
\newcommand{\lblp}{{\decay{\Lb}{\Lz\phi}}\xspace}
\newcommand{\bdkp}{{\decay{\Bd}{\KS\phi}}\xspace}

\usepackage{cite} 
\usepackage{mciteplus}

\usepackage{longtable} 
\usepackage{subfigure}
\begin{document}

\renewcommand{\thefootnote}{\fnsymbol{footnote}}
\setcounter{footnote}{1}


\begin{titlepage}
\pagenumbering{roman}

\vspace*{-1.5cm}
\centerline{\large EUROPEAN ORGANIZATION FOR NUCLEAR RESEARCH (CERN)}
\vspace*{1.5cm}
\noindent
\begin{tabular*}{\linewidth}{lc@{\extracolsep{\fill}}r@{\extracolsep{0pt}}}
\ifthenelse{\boolean{pdflatex}}
{\vspace*{-2.7cm}\mbox{\!\!\!\includegraphics[width=.14\textwidth]{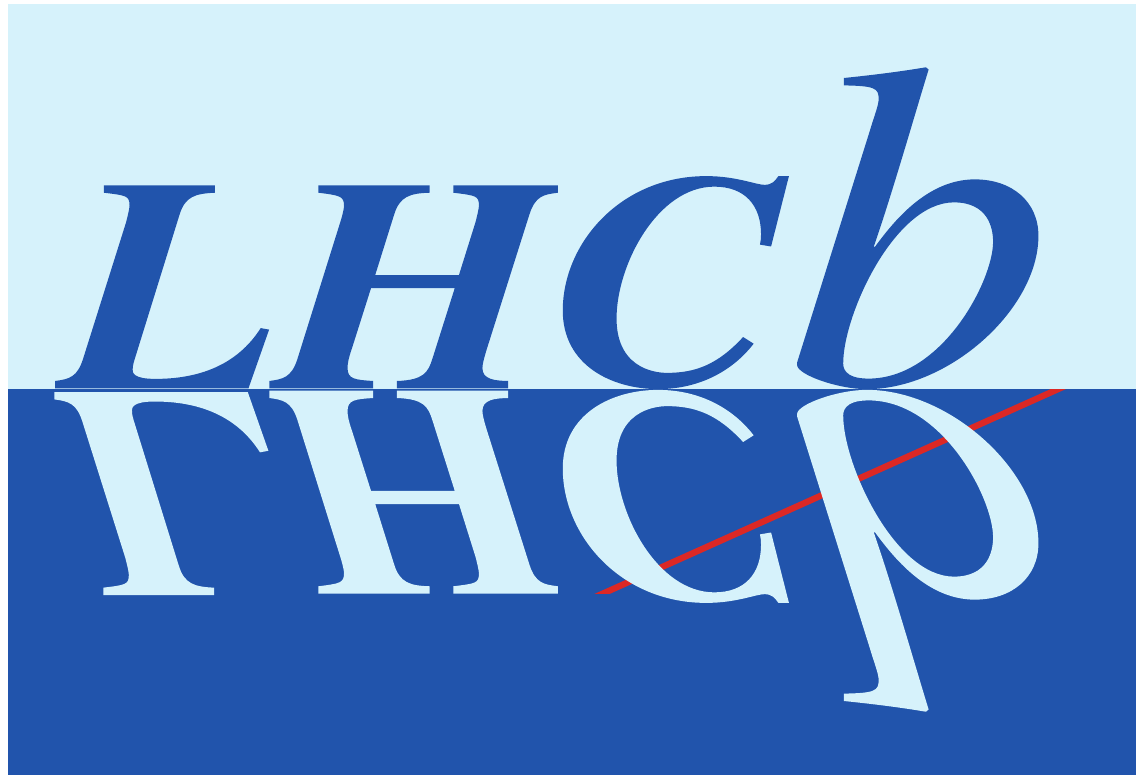}} & &}%
{\vspace*{-1.2cm}\mbox{\!\!\!\includegraphics[width=.12\textwidth]{lhcb-logo.eps}} & &}%
\\
 & & CERN-EP-2016-047 \\  
 & & LHCb-PAPER-2016-002 \\  
 & & June 6, 2016 \\ 
 & & \\
\end{tabular*}

\vspace*{4.0cm}

{\normalfont\bfseries\boldmath\huge
\begin{center}
  Observation of the $\Lb\to\Lz\phi$ decay
\end{center}
}

\vspace*{2.0cm}

\begin{center}
The LHCb collaboration\footnote{Authors are listed at the end of this letter.}
\end{center}

\vspace{\fill}

\begin{abstract}
  \noindent
The $\Lb\to\Lz\phi$ decay is observed using data corresponding to an integrated luminosity of 3.0\invfb
recorded by the \lhcb experiment.
The decay proceeds at leading order via a $\bquark\to\ssbar\squark$ loop transition and is therefore
sensitive to the possible presence of particles beyond the Standard Model. 
A first observation is reported with a significance
of $5.9$ standard deviations. The value of the branching fraction is measured to be
$(5.18\pm1.04\pm0.35\,^{+0.67}_{-0.62})\times10^{-6}$, where the first uncertainty is
statistical, the second is systematic, and the third is related to external inputs.
Triple-product asymmetries are measured to be consistent with zero.
\end{abstract}

\vspace*{2.0cm}

\begin{center}
  Published in Phys.~Lett.~B759 (2016) 282
\end{center}

\vspace{\fill}

{\footnotesize 
\centerline{\copyright~CERN on behalf of the \lhcb collaboration, licence \href{http://creativecommons.org/licenses/by/4.0/}{CC-BY-4.0}.}}
\vspace*{2mm}

\end{titlepage}


\newpage
\setcounter{page}{2}
\mbox{~}
%
%
%
%

\cleardoublepage


\renewcommand{\thefootnote}{\arabic{footnote}}
\setcounter{footnote}{0}



\pagestyle{plain} 
\setcounter{page}{1}
\pagenumbering{arabic}

\section{Introduction}
\label{sec:Introduction}

\begin{figure}[b]
\begin{center}
\includegraphics[height=4cm]{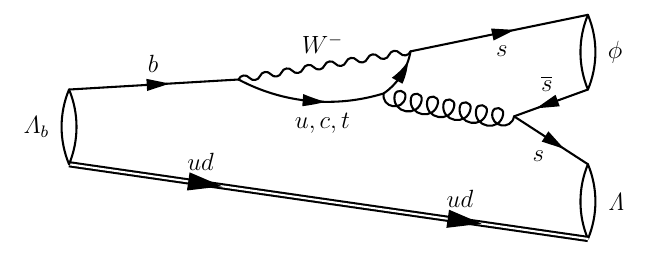}
\end{center}
\caption{\small
Feynman diagram contributing to the $\Lb\to\Lz\phi$ decay.
}
\label{fig:FD}
\end{figure}

In the Standard Model (SM), the flavour-changing neutral current decay 
$\Lb\to\Lz\phi$ proceeds via a $\bquark\to\ssbar\squark$ loop (penguin) process. 
A Feynman diagram of the gluonic penguin that contributes to this decay at leading order
is displayed in Fig.~\ref{fig:FD}. This transition has been the subject of theoretical 
and experimental interest in \Bs and \Bd
decays, since possible beyond the SM particles in the loop could induce
non-SM \CP violation~\cite{Shimizu:2012zw,Datta:2008be,Moroi:2000tk}. 
The process has been probed with decay-time-dependent methods in the $\Bs\to \phi\phi$ and $\bdkp$ 
decay modes~\cite{LHCb-PAPER-2014-026,LHCb-PAPER-2013-007,Abe:2003yt,Lees:2012kxa}, which test for
\CP violation in the interference between mixing and decay.
In addition, measurements of \CP violation in the decay have been performed with the 
flavour-specific \decay{\Bd}{\Kstarz\phi} channel~\cite{LHCb-PAPER-2014-005}.
The results to date are consistent with \CP conservation in the $\bquark\to\ssbar\squark$ process.
Model-independently, non-SM physics contributions could appear differently in these decay modes, though
many models contain strong correlations~\cite{Raidal:2002ph}.

Measurements with \Lb baryons offer the possibility to look for 
\CP violation in the decay, both by studying \CP asymmetries and 
by means of $T$-odd observables. These observables have been
studied in greater detail for \Bs and \Bd meson decays
than those for \Lb baryons~\cite{Gronau:2011cf,Patra:2013xua,LHCb-PAPER-2014-026,LHCb-PAPER-2014-005}. 
Proposed methods to study $T$-odd asymmetries of \Lb baryons~\cite{Leitner:2006sc}
exploit the polarisation structure of $\Lb\to\Lz V$ decays, where $V$ denotes a vector resonance~\cite{Leitner:2006sc},
and can be affected by the initial \Lb polarisation if non-zero.
An \lhcb measurement of the initial polarisation in \decay{\Lb}{\jpsi\Lz} decays has yielded a value consistent with zero, though polarisation at the level of 
$10\%$ is possible given statistical uncertainties~\cite{LHCb-PAPER-2012-057}. 
No SM prediction exists specifically for the $T$-odd asymmetries
in \lblp decays, though no large asymmetries are expected given the prediction of \CP conservation in the decays of beauty mesons for the
same transition.
Measurements of \CP asymmetries have been performed by \lhcb in 
an inclusive analysis of $\Lb\to\Lz hh'$ decays~\cite{DOH}, where $h(h')$ refers to a kaon or pion,
with corresponding \CP asymmetries measured to be
consistent with zero.

In this paper, a measurement of the \lblp branching fraction is presented using
the \bdkp decay as a normalisation channel, which has a measured branching fraction of $(7.3^{+0.7}_{-0.6})\times10^{-6}$~\cite{HFAG}. 
The selection requirements used to isolate the \lblp decay with well-understood efficiencies
reject suitable control channels for a $\Delta A_{\CP}$ measurement.
The \lblp sample is then used to perform
measurements of the $T$-odd triple-product asymmetries, which do not require a control channel.
The results are based on $pp$ collision data
corresponding to an integrated luminosity of  $1.0\invfb$ and $2.0\invfb$ 
collected by the \lhcb experiment at centre-of-mass energies of $\sqrt{s}=7\tev$ in 2011 and 8\tev in 2012, respectively.

\section{Detector and simulation}
\label{sec:Detector}

The \lhcb detector~\cite{Alves:2008zz,LHCb-DP-2014-002} is a single-arm forward
spectrometer covering the \mbox{pseudorapidity} range $2<\eta <5$,
designed for the study of particles containing \bquark or \cquark
quarks. The detector includes a high-precision tracking system
consisting of a silicon-strip vertex detector surrounding the $pp$
interaction region, a large-area silicon-strip detector located
upstream of a dipole magnet with a bending power of about
$4{\mathrm{\,Tm}}$, and three stations of silicon-strip detectors and straw
drift tubes placed downstream of the magnet.
The tracking system provides a measurement of momentum, \ptot, of charged particles with
a relative uncertainty that varies from 0.5\% at low momentum to 1.0\% at 200\gevc.
The minimum distance of a track to a primary vertex, the impact parameter, is measured with a resolution of $(15+29/\pt)\mum$,
where \pt is the component of the momentum transverse to the beam, in\,\gevc.
Different types of charged hadrons are distinguished using information
from two ring-imaging Cherenkov detectors. 
Photons, electrons and hadrons are identified by a calorimeter system consisting of
scintillating-pad and preshower detectors, an electromagnetic
calorimeter and a hadronic calorimeter. 
The online event selection is performed by a trigger, 
which consists of a hardware stage, based on information from the calorimeter and muon
systems, followed by a software stage, which applies a full event
reconstruction.
At the hardware trigger stage, events are required to have a muon with high \pt or a
hadron, photon or electron with high transverse energy in the calorimeters. For hadrons,
the transverse energy threshold is 3.5\gev.
In the subsequent software trigger, at least 
one charged particle must have a transverse momentum $\pt > 1.7$\gevc and be inconsistent with originating from a PV.
Finally, the tracks of two or more of the final-state
particles are required to form a vertex that is significantly
displaced from the PVs. The final state particles that are identified as kaons
are required to have a combined invariant mass consistent with that of the $\phi$ meson.

In the simulation, $pp$ collisions are generated using
\pythiaeight~\cite{Sjostrand:2006za,*Sjostrand:2007gs} 
 with a specific \lhcb
configuration~\cite{LHCb-PROC-2010-056}.  Decays of hadronic particles
are described by \evtgen~\cite{Lange:2001uf}, in which final-state
radiation is generated using \photos~\cite{Golonka:2005pn}. The
interaction of the generated particles with the detector, and its response,
are implemented using the \geant
toolkit~\cite{Allison:2006ve, *Agostinelli:2002hh} as described in
Ref.~\cite{LHCb-PROC-2011-006}.
The decays of \Lb baryons are modelled according to a phase-space description.
Differences in the efficiencies of protons and anti-protons, at the sub-percent level, are accounted for
with the \geant implementation of the detector description.

\section{Selection}
\label{sec:selection}

The $\lblp$ and $\bdkp$ decays are reconstructed through the $\Lz\to\proton\pim$, $\KS\to\pip\pim$ and $\phi\to\Kp\Km$ final states,
where the inclusion of charge conjugate processes is implied throughout the paper.
Decays of \decay{\Lz}{\proton\pim} and \decay{\KS}{\pip\pim}
are reconstructed in two different categories.
The first category contains \Lz (\KS) hadrons that decay inside
the vertex detector acceptance and the
second contains \Lz(\KS) hadrons that decay outside.
These categories are
referred to as \emph{long} and \emph{downstream}, respectively. The high
resolution of the vertex detector leads to enhanced 
momentum, vertex, and mass resolutions for candidates in the long category relative to
downstream candidates.

Boosted decision trees~(BDTs)~\cite{Breiman,AdaBoost} are used to separate signal from
background.
Different BDTs are trained for decays where the daughter tracks of the $\Lz$ (\KS) hadron
are classified as long or downstream and according to whether the data was collected in 2011 (7\tev)
or 2012 (8\tev), yielding eight separate BDTs in total.
The set of input variables used to train the \decay{\Lb}{\Lz\phi}(\decay{\Bd}{\KS\phi}) BDTs
consists of the \Lb (\Bd) vertex fit quality, \pt, $\eta$,
the difference in \chisq of the PV reconstructed with and without the candidate ($\chisq_{\rm IP}$),
the flight distance squared divided by the associated variance ($\chisq_{\rm FD}$),
the angle between the momentum vector and the vector from the PV to the decay vertex,
the $\Lz$ (\KS) vertex fit quality,
and the \pt and $\eta$ of the $\phi$ and
the $\Lz$ (\KS) hadrons. The minimum and maximum values of the \pt and $\eta$ associated to the final state
particles are also included.
In addition, the BDT trained on the long category uses the 
$\chisq_{\rm IP}$ and $\chisq_{\rm FD}$ of the $\Lz$ (\KS) with respect to 
the associated PV. 
A PV is reconstructed by requiring a minimum of five good 
quality tracks that are consistent with originating from the 
same location within the luminous region.
Before the BDTs are trained, initial loose requirements are imposed on the input variables.
The BDTs are trained using simulated candidates for the signal and data sidebands for the background.
For the training samples, the signal
region is defined as being within 150\mevcc of the known \Lb (\Bd) mass~\cite{Agashe:2014kda}. 
In addition, the $\Kp\Km$ invariant mass is required to be within 20\mevcc of the known $\phi$ mass
and the $\proton\pim$ invariant mass is required to be within 15\mevcc of the known \Lz mass~\cite{Agashe:2014kda}.
The sidebands are defined to be within 500\mevcc of the known \Lb (\Bd) mass excluding the signal region.

The figure of merit used to determine the requirement imposed on the \lblp BDT output is 
defined as $\varepsilon/(3/2 + \sqrt{N_{\rm bkg}})$~\cite{Punzi:2003bu}, where $\varepsilon$ is the signal
efficiency, and $N_{\rm bkg}$ is the number of background events. This figure of merit is optimised for detection
at three standard deviations of decay modes not previously observed.
The signal efficiency is obtained from simulated signal candidates
and the number of background events is calculated from fits to the data sidebands 
interpolated to the signal region.
This optimisation procedure is performed separately for each BDT. 

In contrast to the \decay{\Lb}{\Lz\phi} BDTs, the optimum response requirement for the \decay{\Bd}{\KS\phi}
BDTs is chosen based on a figure of merit defined as $N_{\rm sig}/\sqrt{N_{\rm sig}+N_{\rm bkg}}$, where $N_{\rm sig}$ is the number
of signal events, estimated from the BDT efficiency on simulated datasets normalised using the known branching fraction
of the \bdkp decay~\cite{HFAG}, and $N_{\rm bkg}$ is the expected number of background candidates in the signal
region, extrapolated from the \Bd sidebands. This figure of merit is chosen as the \bdkp branching fraction is well measured and 
is optimised separately for each classifier.

\section{Mass fit model}
\label{sec:massmodel}

For both the \decay{\Lb}{\Lz\phi} and \decay{\Bd}{\KS\phi} decay modes, a three-dimensional
fit is employed to determine the signal candidate yields. In the \lblp case, the three dimensions are 
the $\proton\pim\Kp\Km$,
$\proton\pim$, and $\Kp\Km$ invariant masses, while in the fit to determine
the \bdkp candidate yield, the three dimensions are
the $\pip\pim\Kp\Km$,
$\pip\pim$, and $\Kp\Km$ invariant masses.

Four components are present in the \bdkp mass fit:
the signal $\decay{\Bd}{\KS\phi}$ component, the $\Bd\to\KS\Kp\Km$ non-resonant contribution, a \mbox{\BdF} combinatorial 
component, along with a true \KS component combined with two random kaons.
The $\Bd\to\KS\Kp\Km$ non-resonant component has been observed by the \babar~\cite{Aubert:2005ja}, 
\belle~\cite{Abe:2003yt} and \lhcb~\cite{LHCb-PAPER-2013-042} collaborations.
This is separated from the signal decay through the different $\Kp\Km$ invariant mass line shapes.
No significant partially reconstructed background, in which one or more of the final state particles are missed, is found in the \Bd mass
region. 
Peaking backgrounds, from decays in which at least one of the final state particles has been misidentified, 
are suppressed by the narrow $\Kp\Km$ mass window around the $\phi$ meson and are treated as systematic uncertainties.

The \Bd signal is modelled with the same modified Gaussian function as used in Ref.~\cite{LHCB-PAPER-2012-001}. The modified
Gaussian gives extra degrees of freedom to accommodate extended tails far from the mean.
The $\phi$ signal is modelled with a relativistic Breit-Wigner shape~\cite{Abt:2006wt}
convolved with a Gaussian resolution function.
The $\KS$ signal is parametrised by the sum of two Gaussian functions with a common mean.
Decays from real \Bd mesons to the \KS\Kp\Km final state in which the \Kp\Km pair is non-resonant are
described by the same \Bd and \KS line shapes as the signal, but with a phase-space factor to describe the
non-resonant kaon pairs. The phase-space factor is given by the expression $(m^2-(2m_K)^2)/m^2$,
where $m$ is the \Kp\Km invariant mass and $m_K$ is fixed to the value of the charged kaon mass.
The use of a Flatt\'{e} function~\cite{Flatte:1976xv} rather than a phase-space factor to describe a possible scalar component under the $\phi$ resonance
is found to have a negligible effect on the results and is therefore not included.
The combinatorial background is modelled by exponential functions in all three
mass dimensions. 

A simultaneous fit to the {long} and {downstream} datasets is performed.
The \Bd resolution, modified Gaussian tail parameters and resolutions and fractions of the 
\KS Gaussian functions are constrained to values obtained from a fit to simulated data,
performed separateley for {long} and {downstream} datasets.
The total yield and fraction in the downstream dataset are left as free parameters for each component.

The fit to the $\decay{\Lb}{\Lz\phi}$ channel 
uses the same fit model as the \bdkp control channel: a modified Gaussian function is used to describe the \Lb mass shape, 
a double Gaussian model to describe the $\Lz$ shape, and a relativistic Breit-Wigner convolved with a Gaussian resolution function to describe that of the $\phi$ resonance.
Due to the relatively unexplored
mass spectra present in the \lblp decay, the background contributions have been
identified using the data sidebands. In the final fit, four components are present. These are 
the signal $\decay{\Lb}{\Lz\phi}$ component, the $\Lb\to\Lz\Kp\Km$ non-resonant component in which the $\Kp\Km$ dimension
is described using the phase-space factor defined previously,
combinatorial components with true $\phi$ or \Lz resonances, and a component that has a combinatorial origin in all three mass dimensions.
Combinatorial backgrounds are modelled by exponential functions in each fit dimension.
As for the case of the $\decay{\Bd}{\KS\phi}$ fit, the total yield and fraction in the {downstream} dataset 
are left as free parameters for each component. 
In addition, the same parameters are constrained to simulated data as in the
\bdkp fit.

\section{Branching fraction measurement}
\label{sec:BF}

The \lblp branching fraction is obtained from the relation
\begin{align}
\BR(\lblp) = 
\frac{\epsilon^{\rm tot}_{\bdkp}}{\epsilon^{\rm tot}_{\lblp}}
&\cdot\frac{f_{\dquark}}{f_{\Lb}}
\cdot\frac{N_{\lblp}}{N_{\bdkp}}
\cdot\frac{\BR(\Bd\to\Kz\phi)}{2}
\cdot\frac{\BR(\KS\to\pip\pim)}{\BR(\Lz\to\proton\pim)},
\label{eq:BF}
\end{align}
where $\epsilon^{\rm tot}$ denotes the combined efficiency of the candidate reconstruction,
the offline selection, the trigger requirements,
and the efficiency of detector acceptance; $f_{\dquark(\Lb)}$ denotes the fraction of {\bquark} quarks
that hadronise to \Bd (\Lb) hadrons. The ratio is taken from the \lhcb measured value $f_{\Lb}/f_{\dquark}=0.387\pm0.033$~\cite{LHCb-PAPER-2014-004}. 
The extra factor $1/2$ in Eq.~\ref{eq:BF} accounts for the fact
that only half of \Kz mesons will decay as \KS mesons.
The value of the \decay{\Bd}{\Kz\phi} branching fraction is taken to be $(7.3^{+0.7}_{-0.6})\times10^{-6}$~\cite{HFAG},
while the PDG values of the \Lz and \KS branching fractions are used~\cite{Agashe:2014kda}.

The reconstruction, selection and software trigger efficiencies, as well as the acceptance of the \lhcb
detector, are determined from simulated samples, using data-driven correction factors where necessary. 
The different interaction cross-sections of the final-state particles with the detector material
is accounted for using simulated datasets.

For the case of the hardware trigger, the efficiency of events triggered by the signal candidate is 
determined from control samples of \decay{\Dz}{\Km\pip} and \decay{\Lz}{\proton\pim} decays. 
The efficiency of events triggered independently of the signal candidate is determined from simulation.
The agreement between data and simulation for the distributions of the variables used in the BDT 
is verified with the \bdkp data.

Data-driven corrections for the reconstruction efficiency of tracks 
corresponding to the long category are obtained from \jpsi samples using a tag-and-probe method~\cite{Aaij:2014pwa}.
This is applied after a separate weighting to 
ensure agreement in detector occupancy between data and simulation.
For measurements of the relative branching fraction of $\Lb\to\Lz\phi$ to $\Bd\to\KS\phi$, the
final state differs by substituting the proton from the decay of the \Lz with a pion.
However, due to the differences in the kinematics of the pions from the $\Lz$ and the $\KS$ decays,
the distinct correction factors for both daughters of the \Lz and \KS are considered.
In addition to the track reconstruction efficiency, the vertexing efficiency of long-lived
particles contains disagreement between data and simulation.
The corresponding correction factors for the long and downstream datasets are determined separately from $\Dz\to\phi\KS$ decays.

\begin{figure}[t]
\centering
{\includegraphics[width=0.32\textwidth]{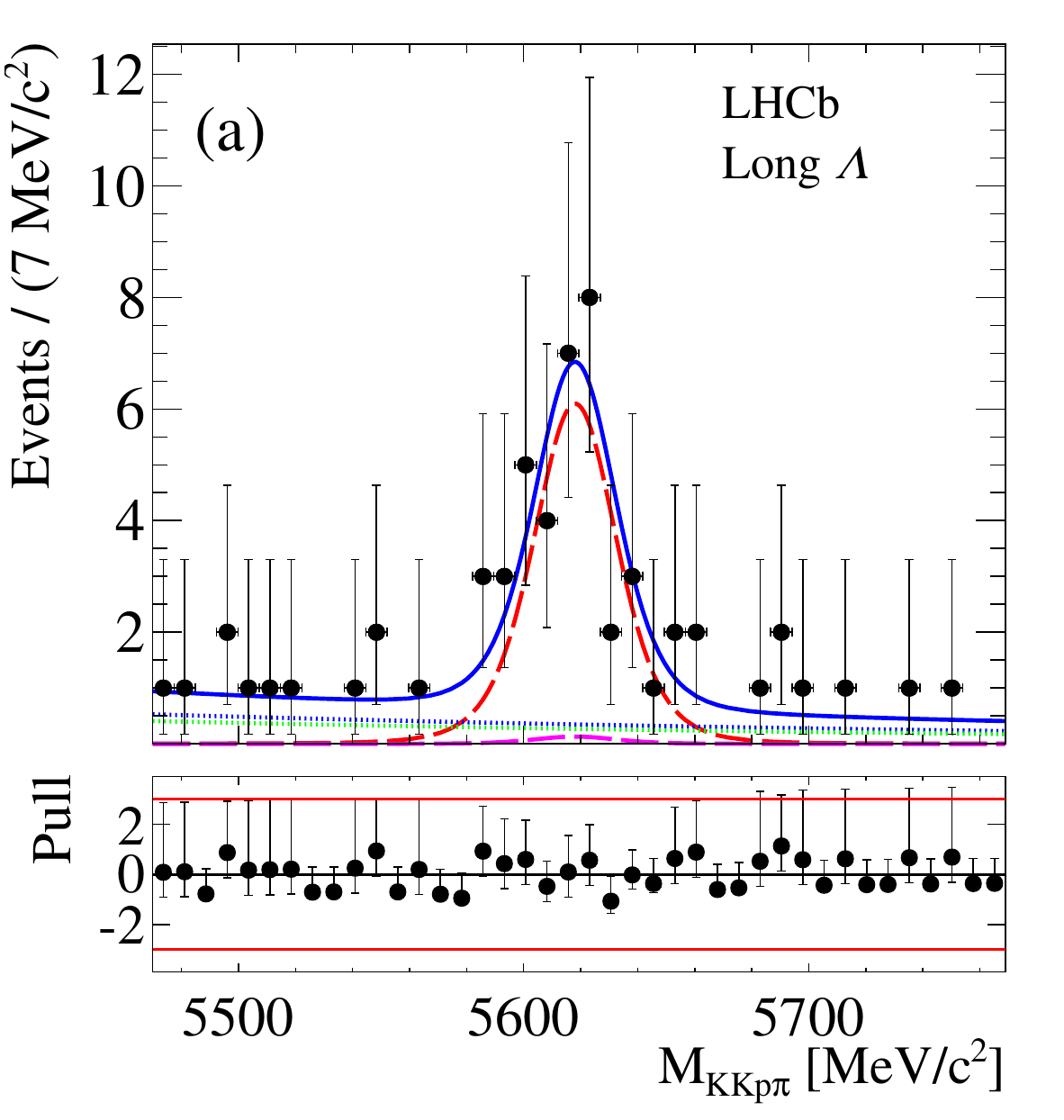}}
{\includegraphics[width=0.32\textwidth]{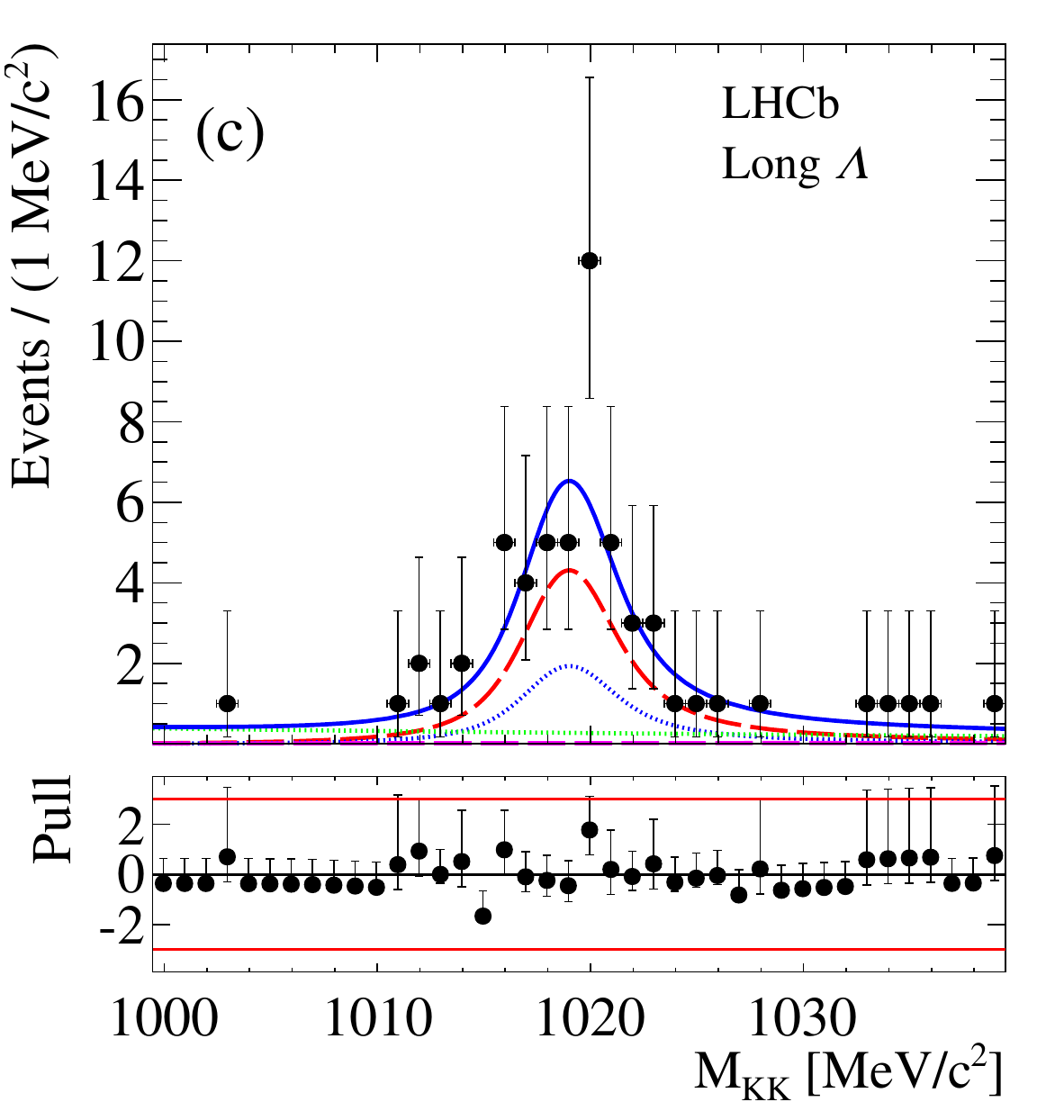}}
{\includegraphics[width=0.32\textwidth]{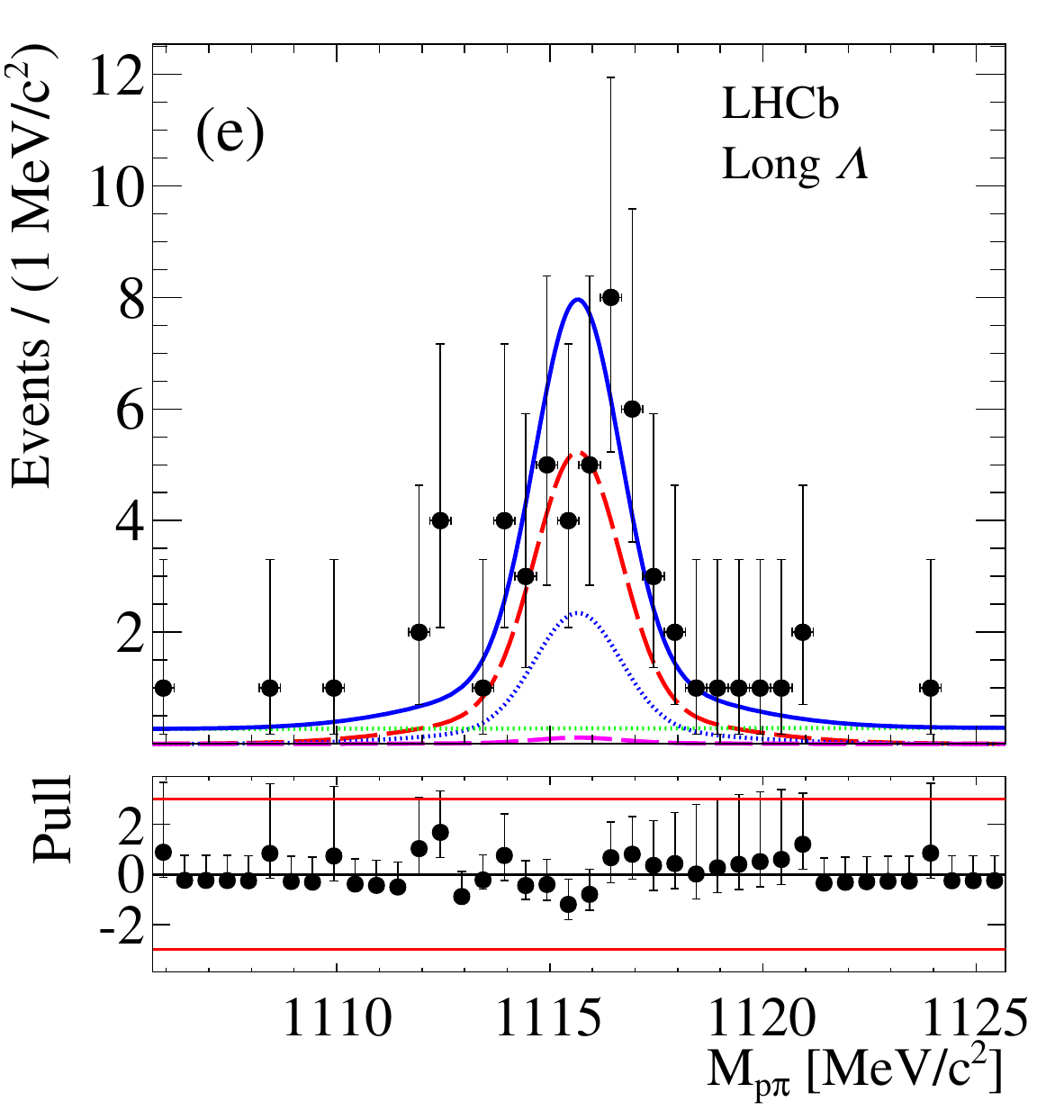}}
{\includegraphics[width=0.32\textwidth]{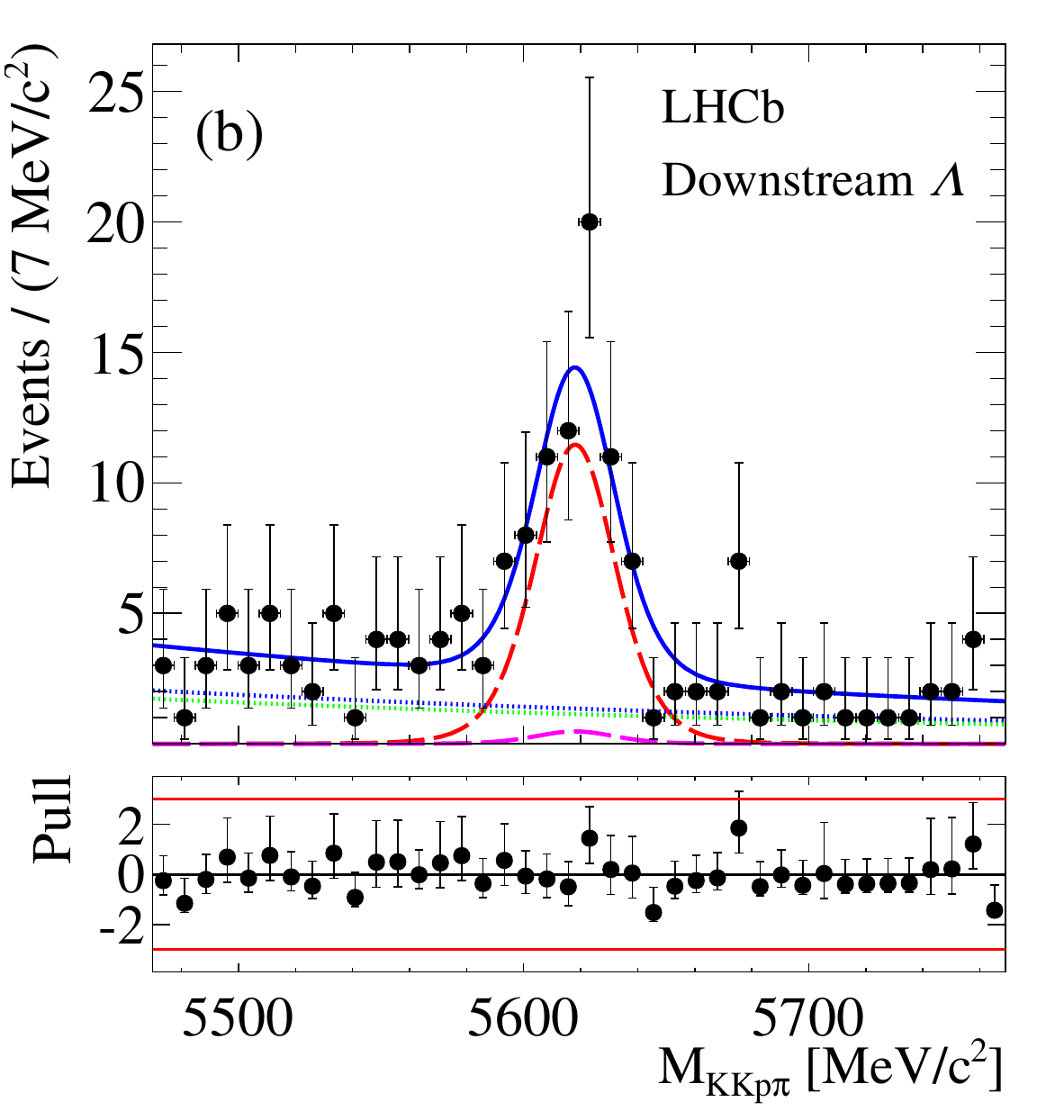}}
{\includegraphics[width=0.32\textwidth]{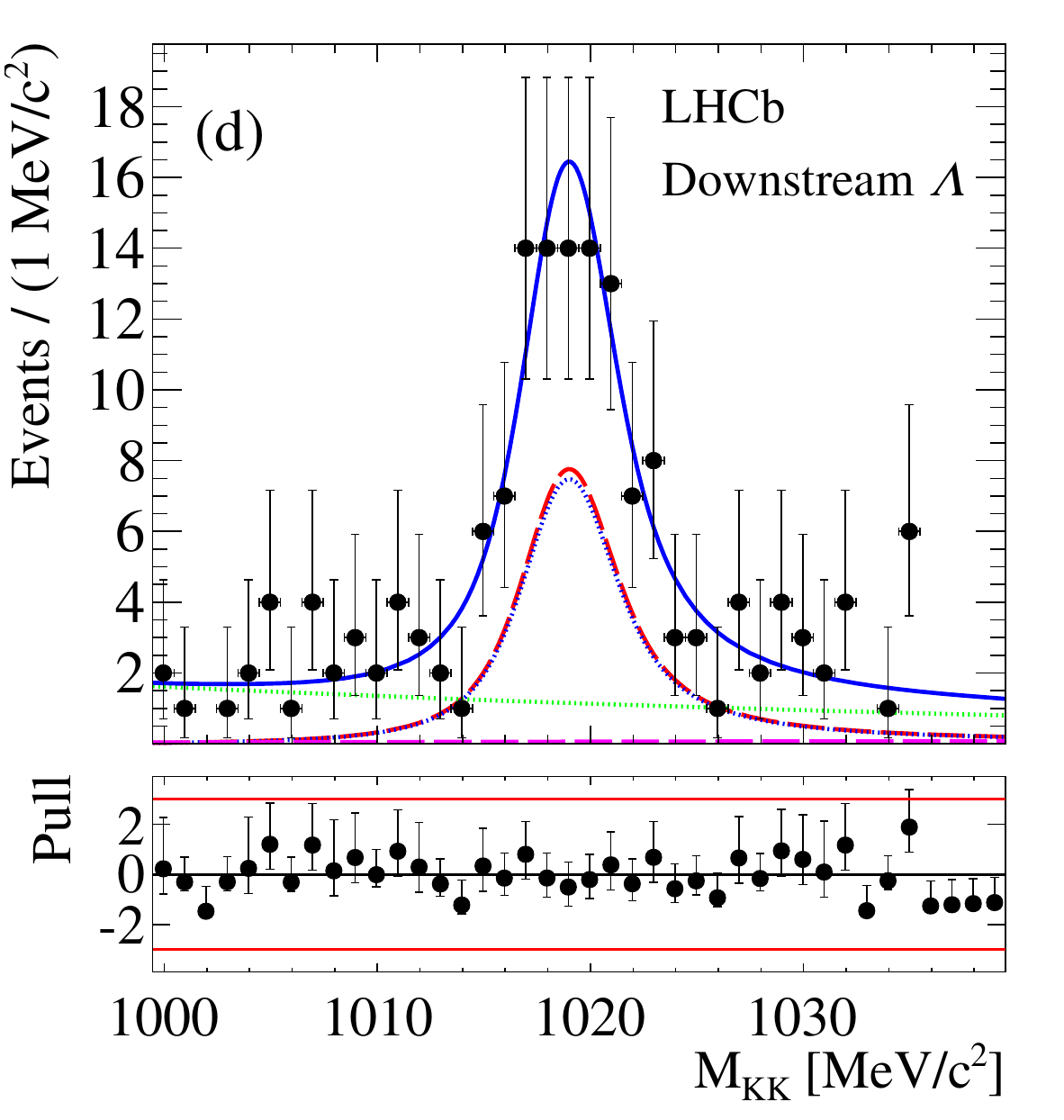}}
{\includegraphics[width=0.32\textwidth]{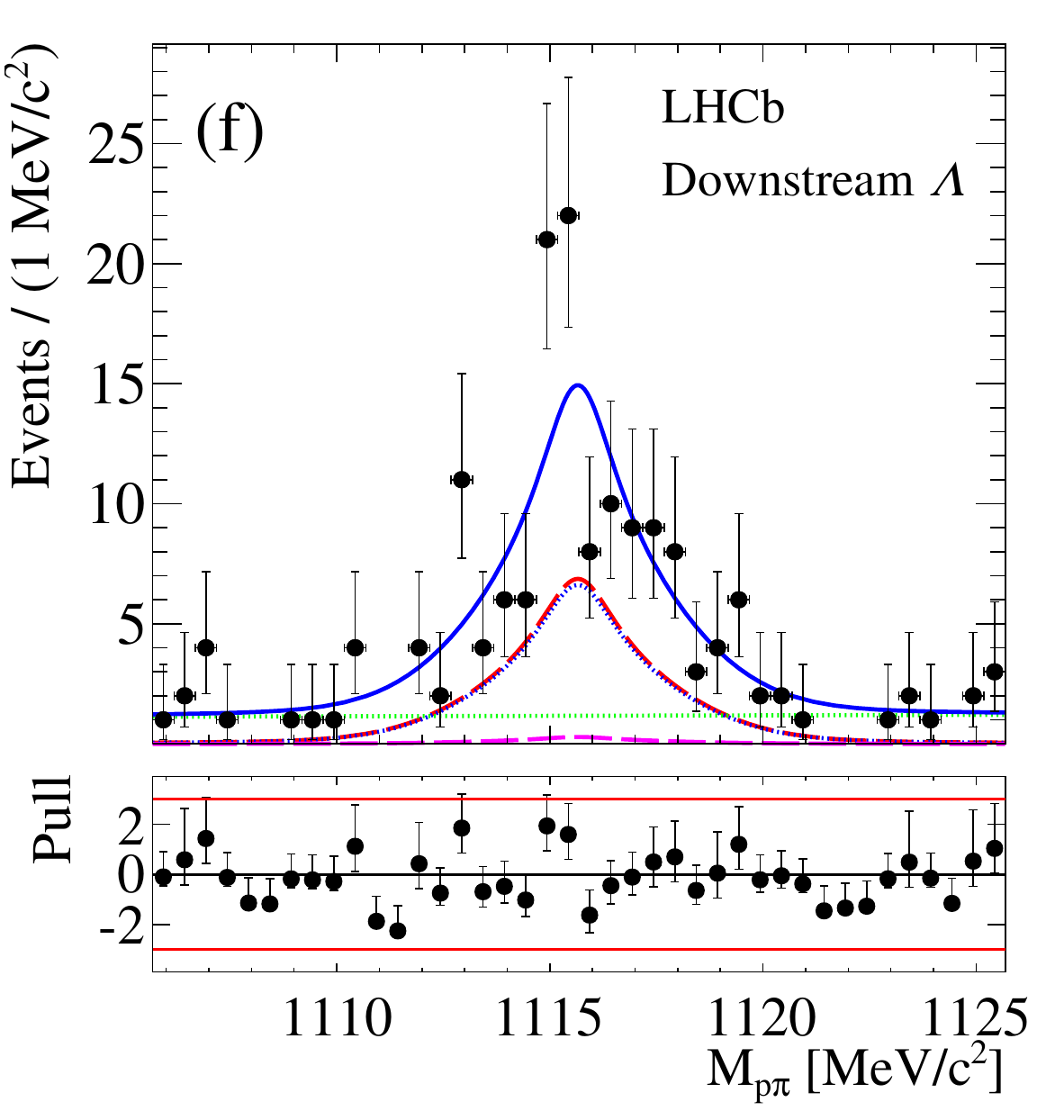}}
\caption{\small
  Fit projections to the \LbF invariant mass in the (a) long and (b) downstream datasets,
  the \phiT invariant mass in the (c) long and (d) downstream datasets, and the \LT invariant mass in the (e) long
  and (f) downstream datasets.
  The total fit projection is given by the blue solid line. The blue and green dotted lines represent
  the $\phi$ + $\Lz$ and pure combinatorial fit components, respectively. The red and magenta dashed lines represent the
  $\lblp$ signal and the $\Lb\to\Lz\Kp\Km$ non-resonant components, respectively.
  Black points represent the data.
  Data uncertainties are Poisson 68\% confidence intervals.
}
\label{fig:LbPhiLam_Data}
\end{figure}
\begin{figure}[t]
\centering
{\includegraphics[width=0.32\textwidth]{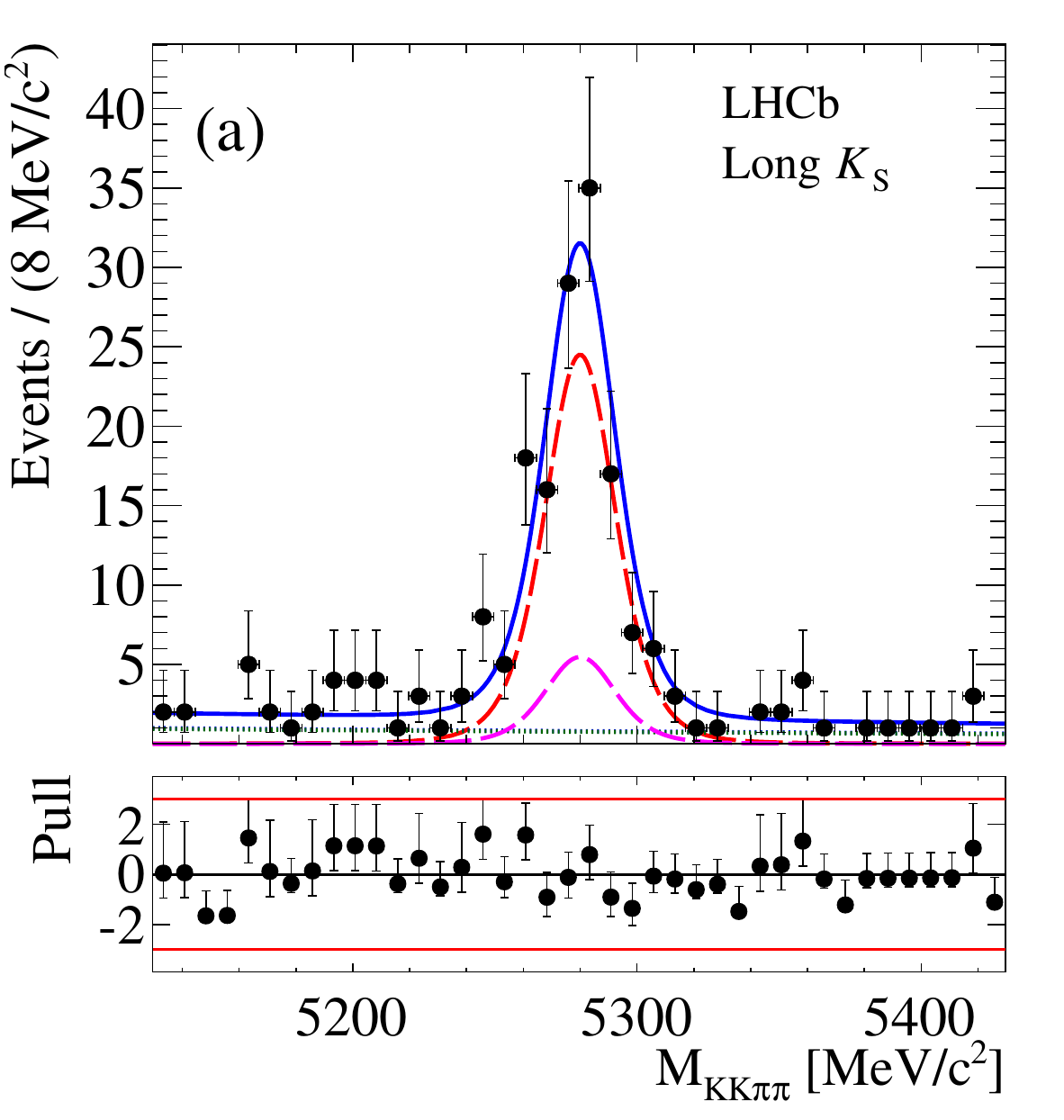}}
{\includegraphics[width=0.32\textwidth]{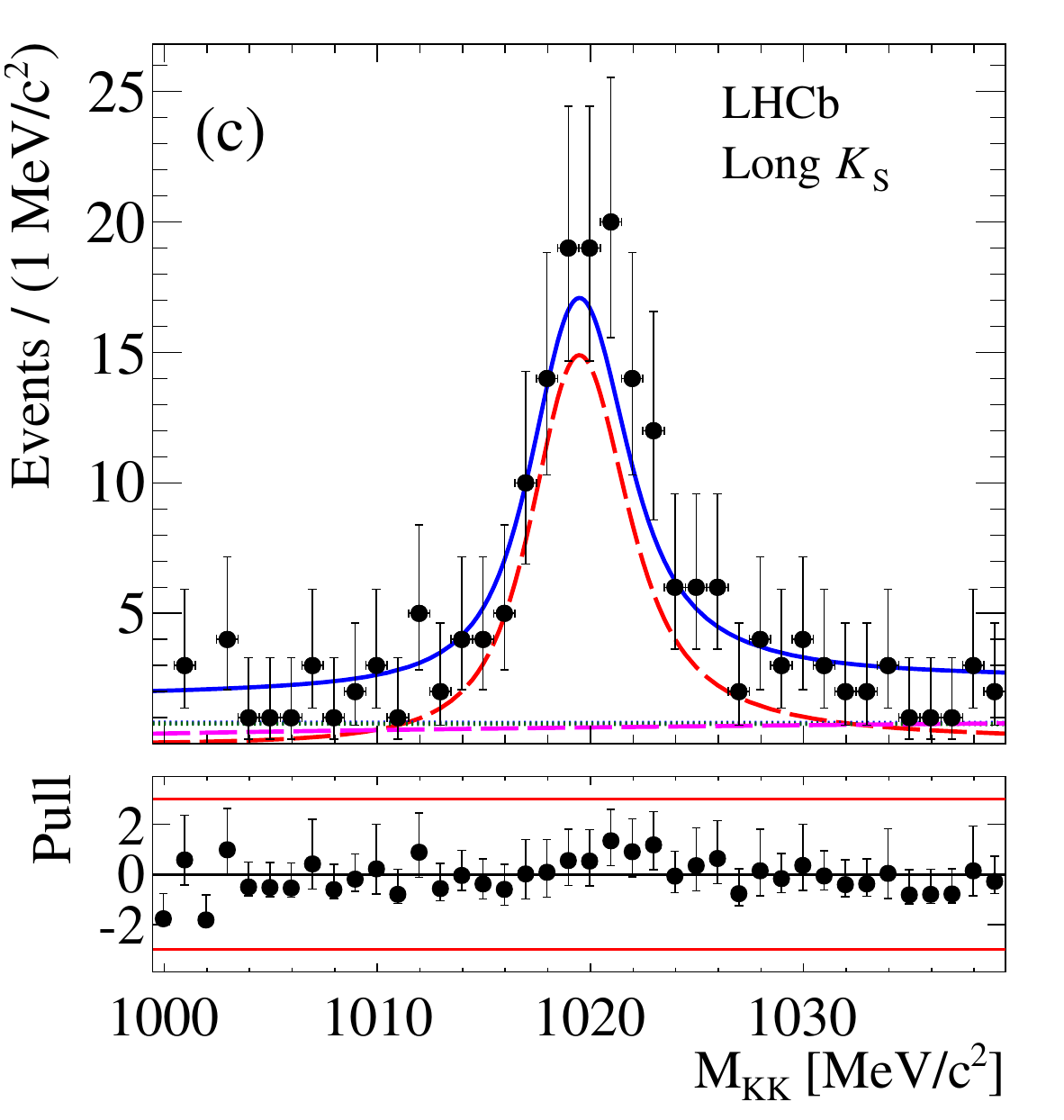}}
{\includegraphics[width=0.32\textwidth]{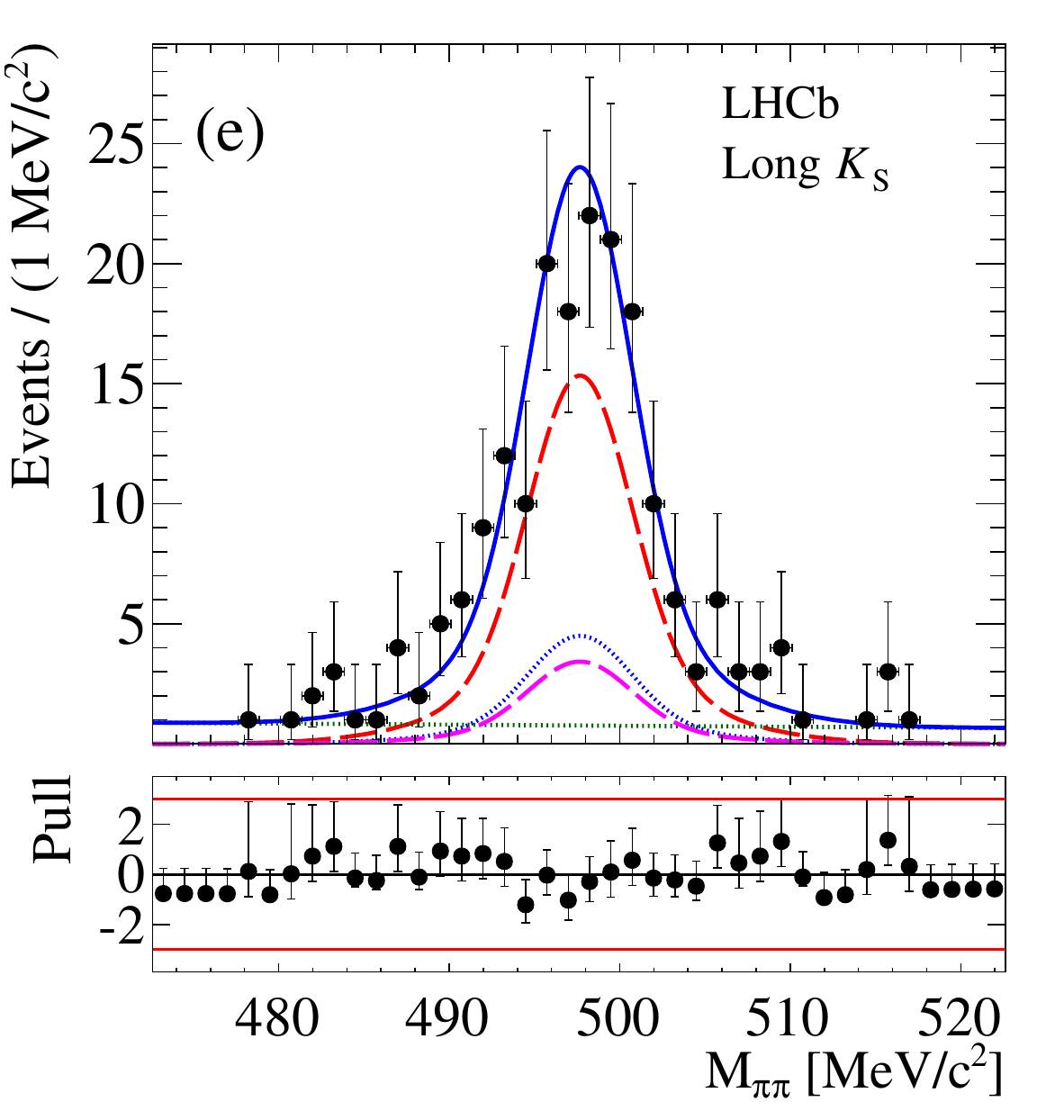}}
{\includegraphics[width=0.32\textwidth]{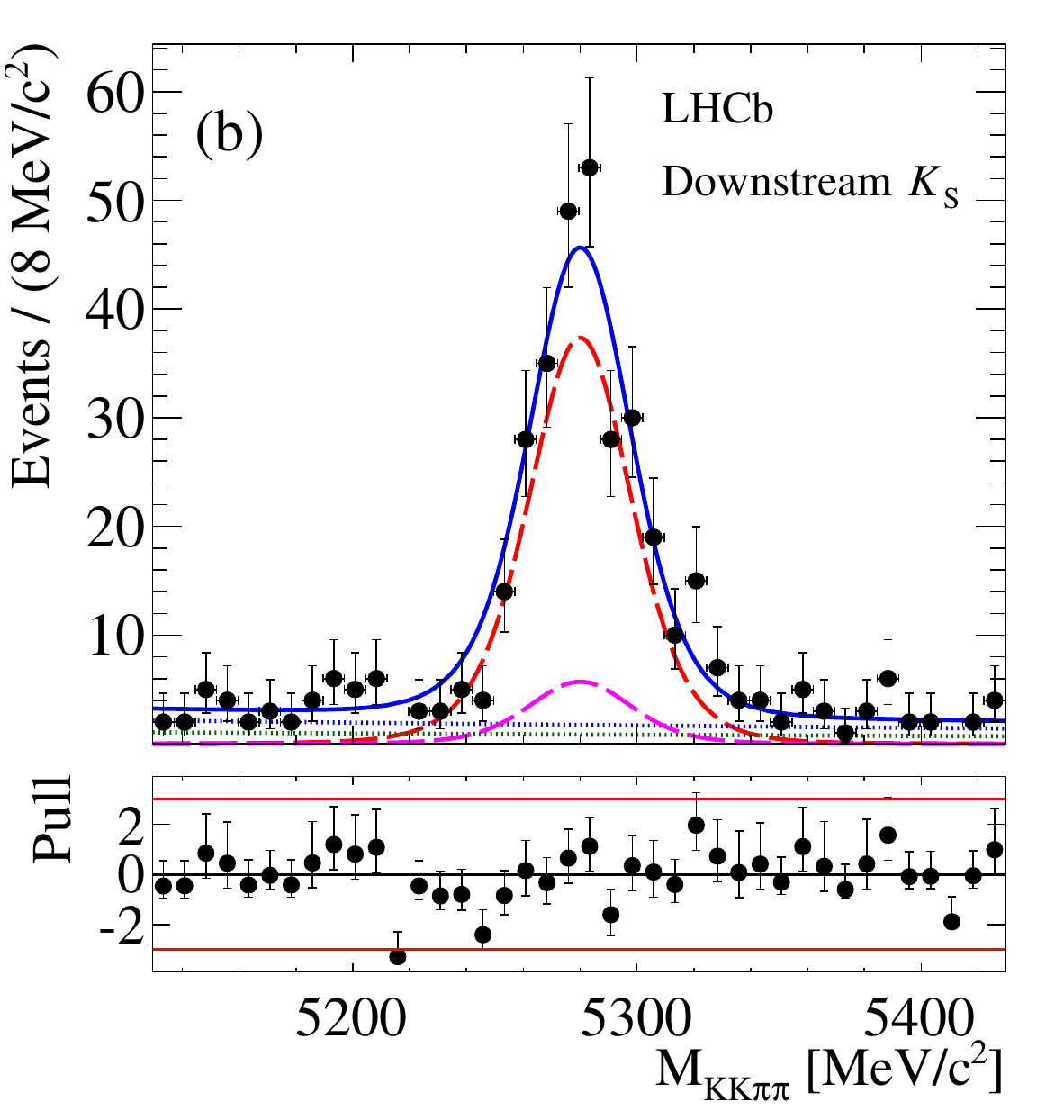}}
{\includegraphics[width=0.32\textwidth]{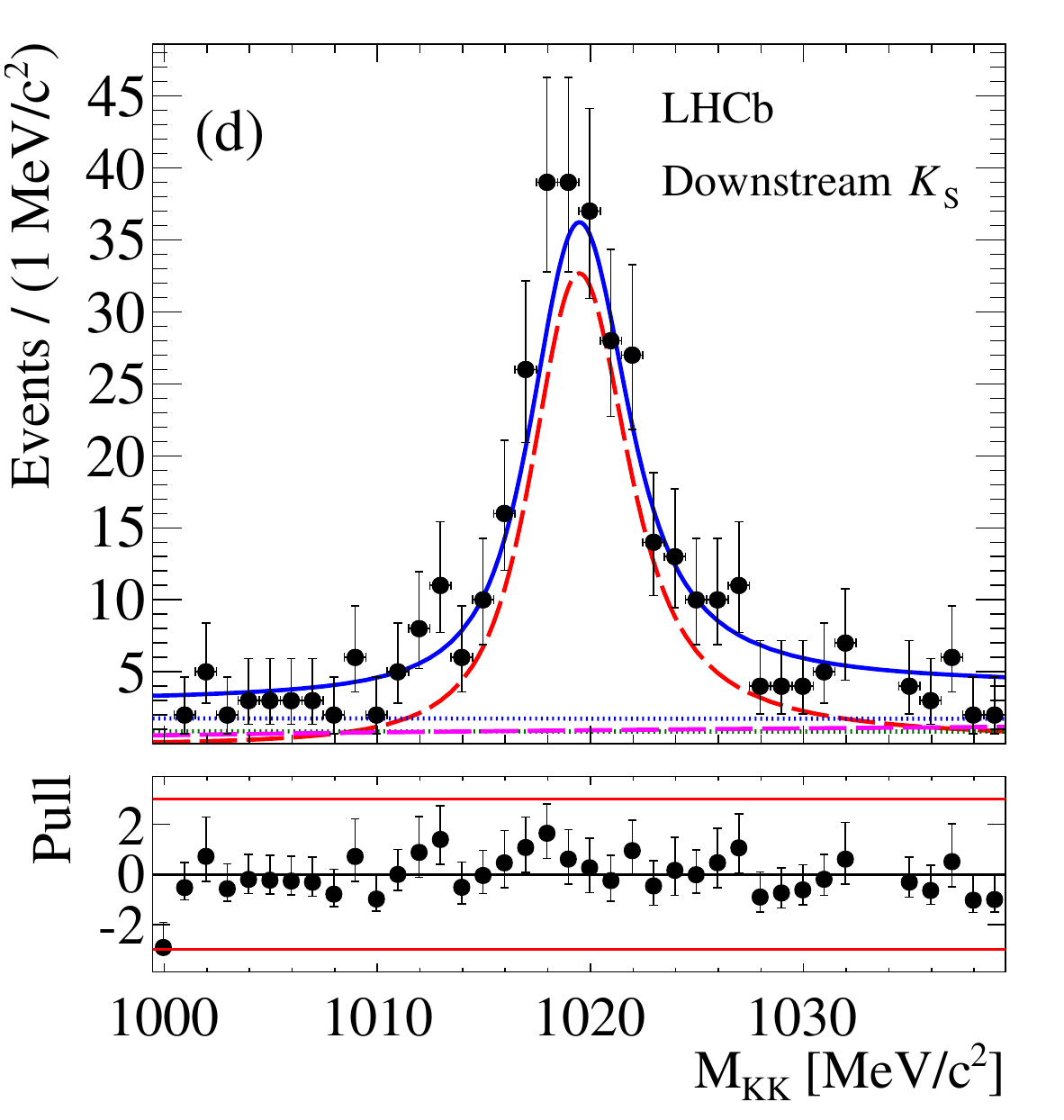}}
{\includegraphics[width=0.32\textwidth]{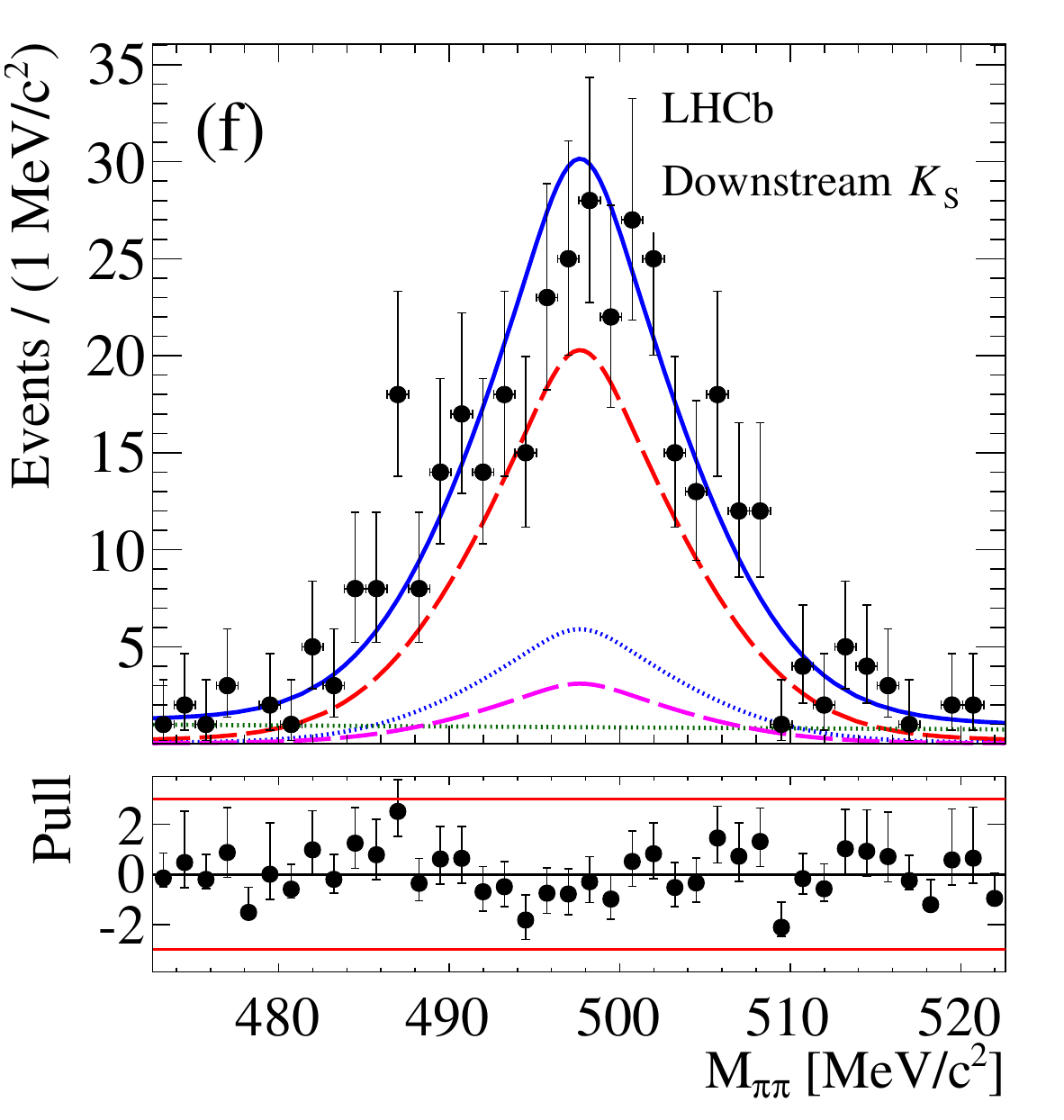}}
\caption{\small
  Fit projections to the \BdF invariant mass in the (a) long and (b) downstream datasets,
  the \phiT invariant mass in the (c) long and (d) downstream datasets, and the {\KST} invariant mass in the (e) long
  and (f) downstream datasets.
  The total fit projection is given by the blue solid line. The green and blue dotted lines represent
  the combinatorial and \KS + random \Kp\Km fit components, respectively. The red and magenta dashed lines represent the
  $\bdkp$ signal and the $\Bd\to\KS\Kp\Km$ non-resonant components, respectively.
  Black points represent the data. 
  Data uncertainties are Poisson 68\% confidence intervals.
}
\label{fig:BdphiKS_Data}
\end{figure}
The yields of the \lblp signal and \bdkp control mode are determined from simultaneous extended
unbinned maximum likelihood fits to the respective
datasets divided according to the data-taking period and also according to whether the $\Lz\,(\KS)$
decay products are reconstructed as long or downstream tracks. Efficiencies are applied to each dataset
individually.
The projections of the fit result to \lblp data are shown in 
Fig.~\ref{fig:LbPhiLam_Data}. 
The fitted yields are $350\pm24$ and $89\pm13$ for the \bdkp and \lblp 
decay modes, respectively. 
The statistical significance of the \lblp decay, determined according to Wilks' theorem~\cite{Wilks:1938dza} 
from the difference in the likelihood value of the fits with and without the \lblp component, is found to be $6.5$
standard deviations.
With the systematic uncertainties discussed below included, the significance of the observed \lblp decay yield is calculated
to be $5.9$ standard deviations. 
The projections of the fit result to the \bdkp data are
shown in Fig.~\ref{fig:BdphiKS_Data}. The fit is found to describe the data well in all three dimensions
and a clear peak from the control mode is seen.

The systematic contributions to the branching fraction uncertainty budget are summarised in Table~\ref{tab:SystSum}.
The largest contributions to the systematic uncertainties result from data-driven corrections
applied to simulated data along with the mass model used to determine the signal yields.
\begin{table}[t]
\caption{\small Systematic uncertainty contributions to the branching fraction ratio.}
\begin{center}
\begin{tabular}{lc}
Source & Uncertainty (\%)\\\hline
Mass model & 3.0 \\
Simulation sample size & 2.2 \\
Tracking efficiency & 0.5\\
Vertex efficiency & 2.6\\
Hardware trigger & 2.8\\
Selection efficiency & 4.1\\
Peaking background & 0.1 \\\hline
Total & 6.7
\end{tabular}
\end{center}
\label{tab:SystSum}
\end{table}

Signal mismodelling is accounted for using
a one-dimensional kernel estimate for the description of the simulated mass distributions~\cite{Cranmer:2000du}.
Background mismodelling is accounted for using a linear function.
The kernel estimate is used in both the signal and control channels to describe the
\Lb, \Bd, \KS, and $\Lz$ line shapes.
In order to determine the systematic uncertainties, 1000 pseudoexperiments are generated 
with the alternative model and are subsequently fitted with the nominal model. The average difference
between the generated and fitted yield values 
is taken as the systematic uncertainty. 
This leads to uncertainties of 3.0\% and 0.6\% for the signal and control mode yields, respectively. 

Systematic uncertainties associated with the efficiency corrections from simulated datasets are considered.
The limited size of the simulated sample gives rise to an uncertainty of 2.2\%.
The main uncertainties in the tracking and vertexing correction factors arise from the limited size of the control sample,
which leads to uncertainties of 0.5\% and 2.6\%, respectively. 
For the case of the trigger efficiency, uncertainties related to the software trigger cancel between the signal and control
modes, as the software trigger decision is made only on the decay products of the $\phi$ meson.
Uncertainties in the efficiency of the hardware trigger selections are estimated using data-driven methods, 
for which an uncertainty of 2.8\% is applied.
The BDTs used to select signal and control modes use the same input variables.
Biases could exist if the simulation mismodels these variables differently for signal and control modes.
In order to quantify this effect, the control mode is selected with the same classifier as the signal decay. The difference
in the measured branching fraction is found to be 4.1\%.

The $\Lb\to\PSigma^0(\to\Lz\gamma)\Kp\Km$
and $\Lb\to\proton\Km\phi$ decay modes are found to be the only significant peaking background contributions. 
However, for the case of the $\Lb\to\proton\Km\phi$ decay, the resulting candidates are reconstructed in
the {long} dataset only. With the assumption that the branching fraction for this decay is the same size as for the signal,
the contribution is $<1\%$ compared to the \lblp decay and far from the \Lb signal region, and is therefore ignored.
In order to determine the shape in the $\proton\pim\Kp\Km$ spectrum of the $\Lb\to\PSigma^0\Kp\Km$ decay, a sample of $\Lb\to\PSigma^0\Kp\Km$ simulated events is used
with a requirement that the $\Kp\Km$ invariant mass is within 30\mevcc of the nominal $\phi$ mass. 
The inclusion of an additional fit component using the shape from simulation is found to have a small effect
on the signal yield at the level of 0.1\%, which is assigned as a systematic uncertainty.
For the case of the $\bdkp$ control mode, no peaking background contributions have been identified.

The branching fraction ratio is measured to be
\begin{align}
\frac{\BR(\lblp)}{\BR(\bdkp)}\frac{f_{\Lb}}{f_{\Bd}} = 0.55\pm 0.11\stat \pm 0.04\syst. \nonumber
\label{eq:BRratio}
\end{align}
The use of the world average value of $\BR(\bdkp)=(3.65\,^{+0.35}_{-0.30})\times10^{-6}$~\cite{HFAG}
gives the final result of
\begin{align}
\BR(\lblp) / 10^{-6}&=\nonumber \\ 
&5.18\pm1.04\stat\pm0.35\syst\,^{+0.50}_{-0.43}\,(\BR(\bdkp))\pm0.44\,(f_d/f_{\Lb}). \nonumber
\end{align}

\section{Triple-product asymmetries}
\label{sec:triple}

\begin{figure}[b]
\begin{center}
\includegraphics[width=0.7\textwidth]{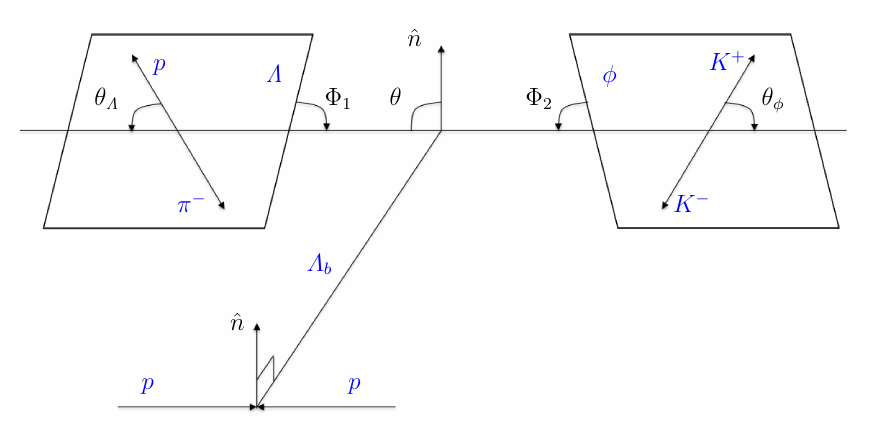}
\end{center}
\caption{\small
  Decay angles for the \lblp decay, where the angles are defined in the text.
}
\label{fig:angles}
\end{figure}

The \lblp decay is a spin-$1/2$ to spin-$1/2$ plus vector transition.
Five angles are needed to describe this decay since \Lb baryons may potentially be produced with a 
transverse polarisation in proton-proton collisions~\cite{LHCb-PAPER-2012-057}, as shown in Fig.~\ref{fig:angles}.
The angle $\theta$ is defined as the
polar angle of the $\Lz$ baryon in the \Lb rest frame with respect to the
normal vector defined through
\begin{align}
\hat{n} = \frac{ \vec{p}_1 \times \vec{p}_{\Lb} }{| \vec{p}_1 \times \vec{p}_{\Lb} |},
\end{align}
where $\vec{p}_1$ is the momentum of an incoming proton and $\vec{p}_{\Lb}$ is the momentum
of the \Lb baryon. 
The angles $\theta_\Lz$ and $\Phi_\Lz$ are defined as the polar and azimuthal
angles of the proton from the decay of the $\Lz$ baryon in the $\Lz$ rest frame.
The angles $\theta_\phi$ and $\Phi_\phi$ are defined as the polar and azimuthal angles of the \Kp meson
in the rest frame of the $\phi$ meson.

Triple-product asymmetries, which are odd under time-reversal, 
have been proposed by Leitner and Ajaltouni using the azimuthal angles $\Phi_{n_i},i\in\{\Lz,\phi\}$,
defined as~\cite{Leitner:2006sc}
\begin{align}
\cos\Phi_{n_i} &= \vec{e}_Y\cdot\vec{u}_i,\\
\sin\Phi_{n_i} &= \vec{e}_Z\cdot(\vec{e}_Y\times\vec{u}_i),
\end{align}
where
\begin{align}
\vec{u}_i = \frac{\vec{e}_Z\times\hat{n}_i}{|\vec{e}_Z\times\hat{n}_i|}.
\end{align}
The basis $\{\vec{e}_X,\vec{e}_Y,\vec{e}_Z\}$ is defined in the \Lb rest frame,
in which $\vec{e}_Z$ is parallel to $\hat{n}$, $\vec{e}_X$ is chosen to be parallel to
the momentum of the incoming proton, and $\hat{n}_{\Lz(\phi)}$ is the
normal vector to the $\Lz(\phi)$ decay plane, defined through
\begin{align}
\hat{n}_\Lz &= \frac{\vec{p}_p\times\vec{p}_\pi}{|\vec{p}_p\times\vec{p}_\pi|},\\
\hat{n}_\phi &= \frac{\vec{p}_{\Kp}\times\vec{p}_{\Km}}{|\vec{p}_{\Kp}\times\vec{p}_{\Km}|}.
\end{align}
Asymmetries in $\cos\Phi_{n_i}$ and $\sin\Phi_{n_i}$, where $i\in\{\Lz,\phi\}$, are defined as
\begin{align}
A_{i}^c &= \frac{N^{+,c}_i - N^{-,c}_i}{N^{+,c}_i + N^{-,c}_i},\\
A_{i}^s &= \frac{N^{+,s}_i - N^{-,s}_i}{N^{+,s}_i + N^{-,s}_i},
\end{align}
where $N^{+(-),c}_i$ and $N^{+(-),s}_i$ denote the number of candidates for which the
$\cos\Phi_{n_i}$ and $\sin\Phi_{n_i}$ observables are positive (negative), respectively.

The asymmetries $A_\Lz^{c,s}$ and $A_\phi^{c,s}$ are determined experimentally through a simultaneous 
unbinned maximum likelihood fit to datasets in which the
relevant observables are positive and negative.
The fit construction and observables are identical to that used for the branching fraction measurement.
However, the yields for each dataset are parametrised in terms of the total yield, $N_j$, and
the asymmetry, $A_j$, for fit component $j$ as
\begin{align}
N^+_j &= \frac{N^j}{2}(1+A_j), \\
N^-_j &= \frac{N^j}{2}(1-A_j).
\end{align}

\begin{figure}[t]
\centering
{\includegraphics[width=0.4\textwidth]{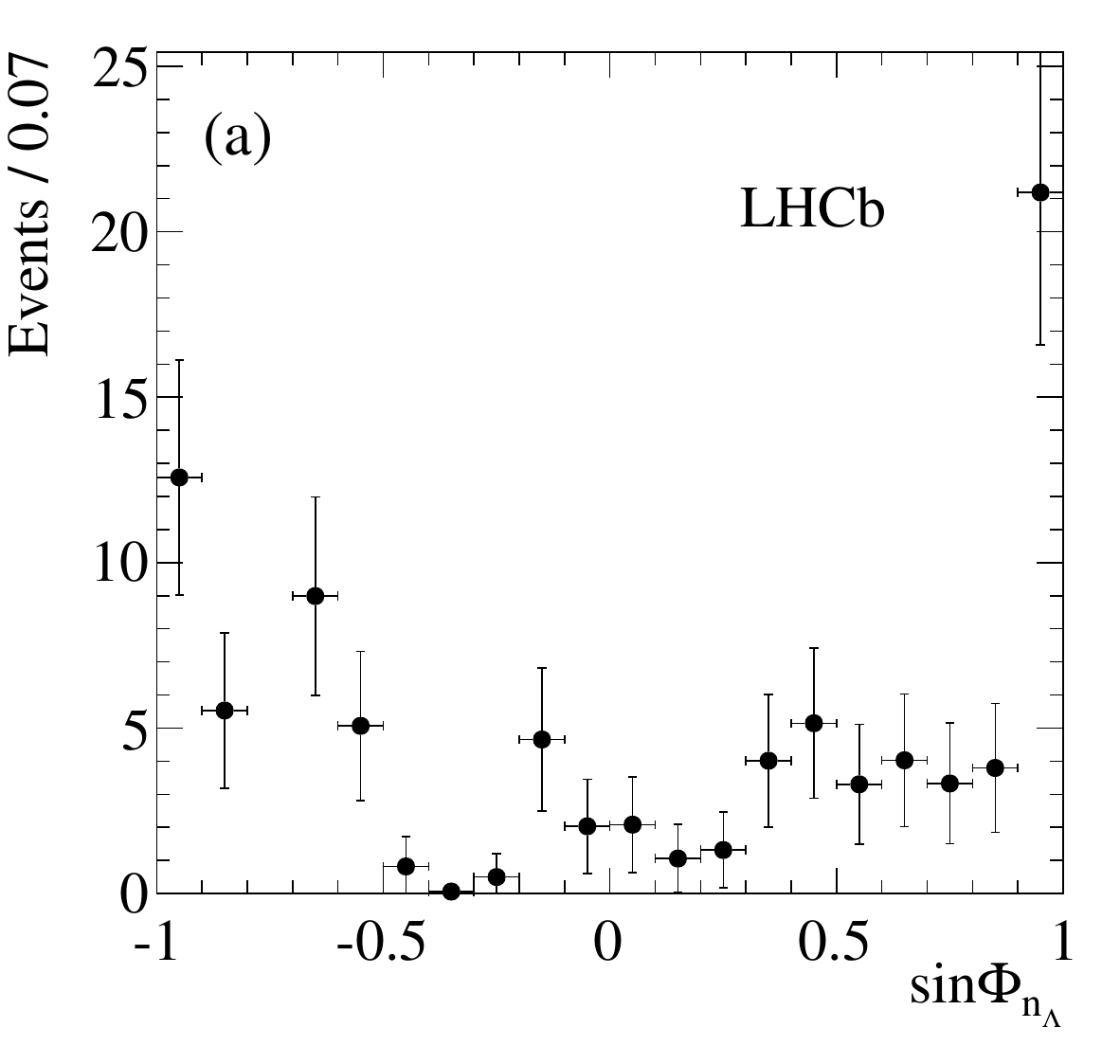}}
{\includegraphics[width=0.4\textwidth]{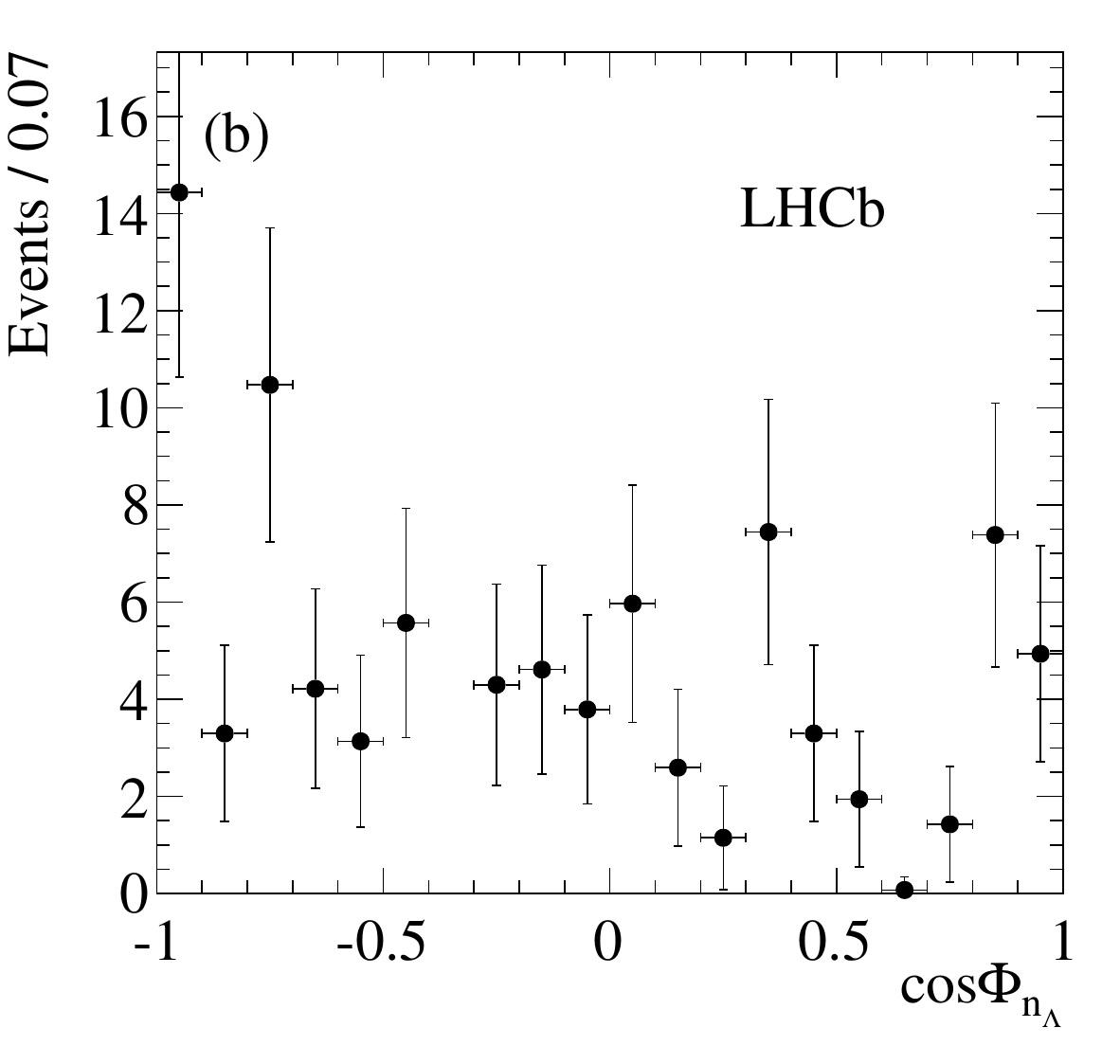}}
{\includegraphics[width=0.4\textwidth]{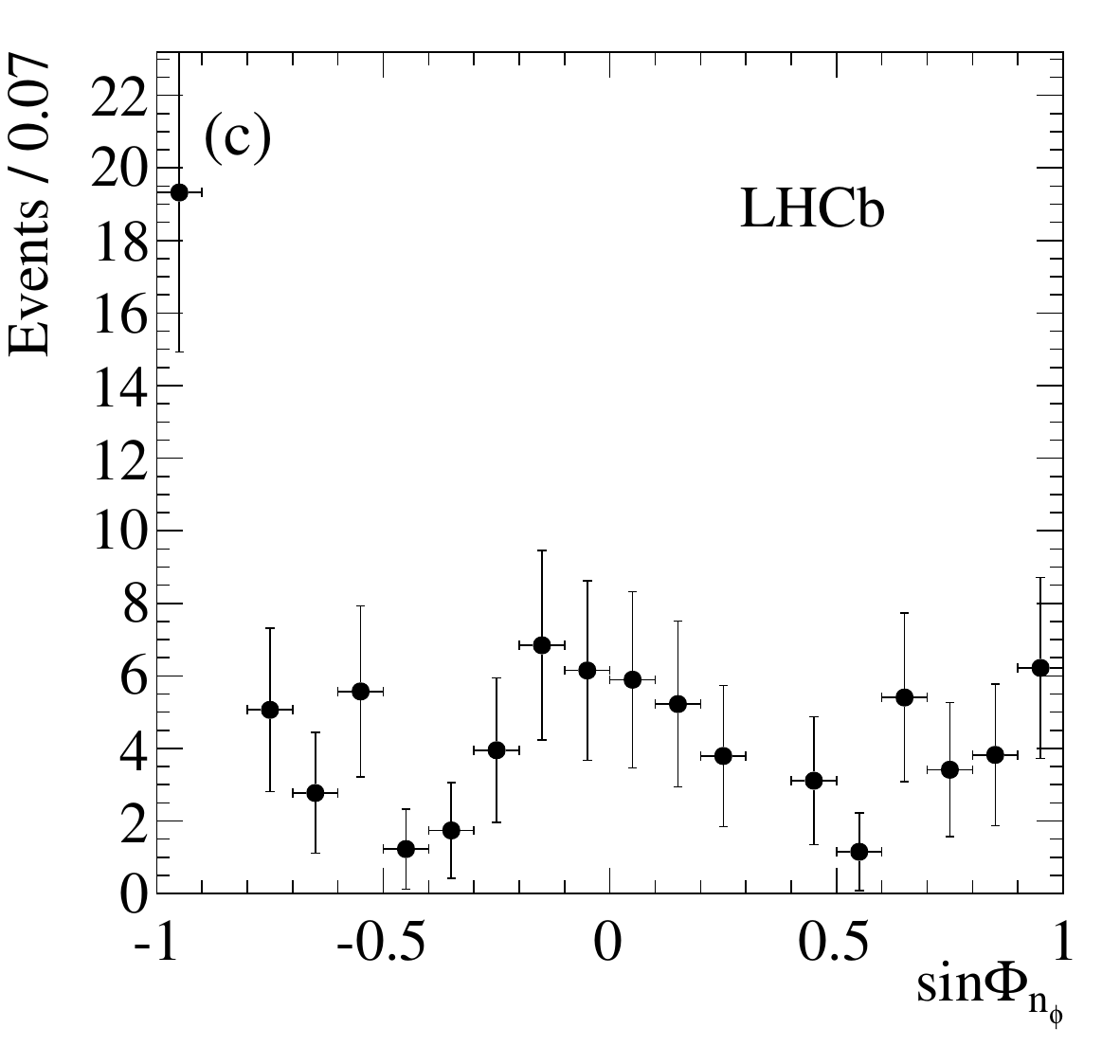}}
{\includegraphics[width=0.4\textwidth]{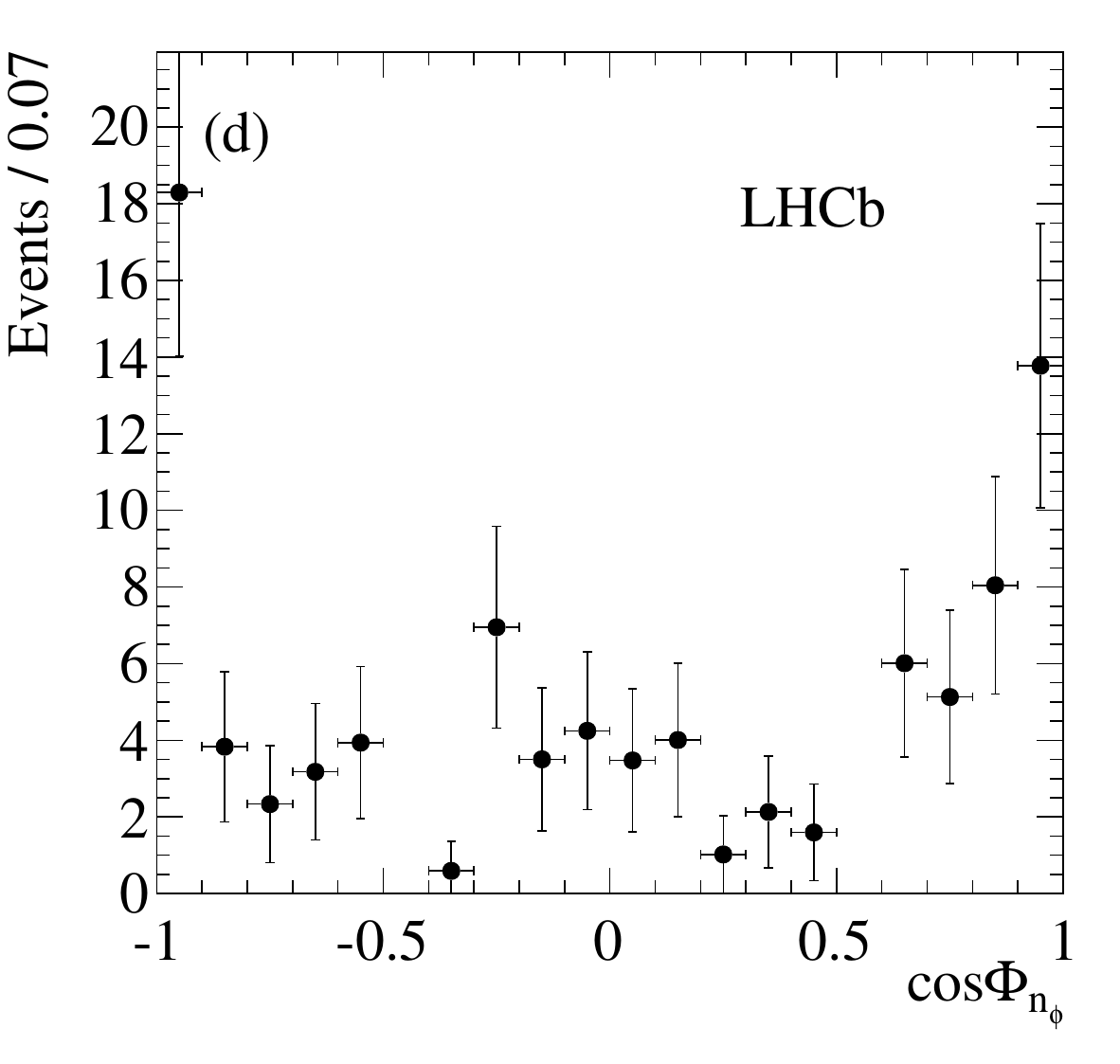}}
\caption{\small Distributions of the angular observables: (a) $\sin\Phi_{n_\Lz}$, 
(b) $\cos\Phi_{n_\Lz}$, (c) $\sin\Phi_{n_\phi}$, (d) $\cos\Phi_{n_\phi}$ from 
weighted \lblp data.}
\label{fig:angobsData}
\end{figure}

Distributions of the $\sin\Phi_{n_{(\Lz,\phi)}}$ and $\cos\Phi_{n_{(\Lz,\phi)}}$ observables
from \lblp data have been extracted using the \sPlot method~\cite{Pivk:2004ty} and
are provided in Fig.~\ref{fig:angobsData}. The numerical values of the fitted asymmetries
are given in Table~\ref{tab:simasymData}.
\begin{table}[b]
\caption{\small Asymmetries measured from \lblp data events.}
\centering
\begin{tabular}{lr}
Asymmetry & Fit value \\ \hline
$A^c_\Lz$ & $-0.22\pm0.12$\\
$A^s_\Lz$ & $0.13\pm0.12$\\
$A^c_\phi$ & $-0.01\pm0.12$\\
$A^s_\phi$ & $-0.07\pm0.12$\\
\end{tabular}
\label{tab:simasymData}
\end{table}

Mismodelling of the mass components could lead to background contamination
in the determination of the asymmetries. In the determination of the uncertainty related to the mass model, two contributions are considered.
These are the line shape models and the background asymmetries.
The effects of the line shapes are quantified using the same method as the branching fraction
measurement, \ie the generation of datasets with 
a one-dimensional kernel estimate of the simulation mass distributions in addition
to modification of the background description.
In the nominal fit, components that are not from the \lblp signal 
have zero asymmetries. For background components this is justified due to the uncorrelated
kinematics of the \Kp\Km and \proton\pim systems. However, the non-resonant $\Lb\to\Lz\Kp\Km$
contribution could have non-zero asymmetries.
The systematic uncertainty due to the assumption
of zero background asymmetries is determined through comparing the nominal fit against
the fit with all possible asymmetries allowed to vary freely.

Efficiencies are found to be independent of the $\sin\Phi_{n_i}$ and $\cos\Phi_{n_i}$ observables. 
The systematic uncertainty due to the angular acceptance is then taken from the 
statistical uncertainty in fits to the simulated datasets,
after the application of an appropriate weighting to account for the differences between data and simulation.
The resolutions of the angular observables are found from simulated events to be 
32.3\mrad and 22.1\mrad for the $\Phi_{n_\Lz}$ and $\Phi_{n_\phi}$ angles, respectively. 
The uncertainty due to bin migration is then assigned assuming maximal asymmetry and leads to minor uncertainties of 
0.007 for the $\Phi_{n_\phi}$ angle and 
0.010 for the $\Phi_{n_\Lz}$ angle. 
Systematic contributions to the triple-product uncertainty  budget are summarised in Table~\ref{tab:SystSum_triple}.
\begin{table}[t]
\caption{\small Systematic uncertainty contributions to the triple-product asymmetries.}
\begin{center}
\begin{tabular}{lrrrr}
Source & $A_\Lz^c$  & $A_\Lz^s$ & $A_\phi^c$  & $A_\phi^s$ \\\hline
Mass model         & 0.061 & 0.051 & 0.026 & 0.009\\
Angular acceptance & 0.010 & 0.010 & 0.010 & 0.010 \\
Angular resolution & 0.008 & 0.008 & 0.005 & 0.005  \\\hline
Total              & 0.062 & 0.053 & 0.028 & 0.014
\end{tabular}
\end{center}
\label{tab:SystSum_triple}
\end{table}

\section{Summary}
\label{sec:summary}

A search for the $\Lb\to\Lz\phi$ decay is presented based on a dataset of 3.0\invfb collected
by the \lhcb experiment in 2011 and 2012.
The decay is observed for the first time with a significance
of $5.9$ standard deviations including systematic uncertainties. The branching fraction is found to be
\begin{align}
\BR(\lblp) / 10^{-6}&=\nonumber \\ 
&5.18\pm1.04\stat\pm0.35\syst\,^{+0.50}_{-0.43}\,(\BR(\bdkp))\pm0.44\,(f_d/f_{\Lb}). \nonumber
\end{align}
Triple-product asymmetries are measured to be
\begin{align}
A^c_\Lz &=-0.22\pm0.12\stat\pm0.06\syst, \nonumber\\
A^s_\Lz &=\phantom{-}0.13\pm0.12\stat\pm0.05\syst, \nonumber\\
A^c_\phi &=-0.01\pm0.12\stat\pm0.03\syst,   \nonumber\\
A^s_\phi &=-0.07\pm0.12\stat\pm0.01\syst \nonumber,
\end{align}
and are consistent with zero.
Data collected by the \lhcb experiment in the forthcoming years will improve the statistical precision of these measurements and enable
the dynamics of $\bquark\to\squark$ transitions in beauty baryons 
to be probed in greater detail, which will greatly enhance the reach of searches for physics beyond the 
SM.

\section*{Acknowledgements}

\noindent We express our gratitude to our colleagues in the CERN
accelerator departments for the excellent performance of the LHC. We
thank the technical and administrative staff at the LHCb
institutes. We acknowledge support from CERN and from the national
agencies: CAPES, CNPq, FAPERJ and FINEP (Brazil); NSFC (China);
CNRS/IN2P3 (France); BMBF, DFG and MPG (Germany); INFN (Italy); 
FOM and NWO (The Netherlands); MNiSW and NCN (Poland); MEN/IFA (Romania); 
MinES and FANO (Russia); MinECo (Spain); SNSF and SER (Switzerland); 
NASU (Ukraine); STFC (United Kingdom); NSF (USA).
We acknowledge the computing resources that are provided by CERN, IN2P3 (France), KIT and DESY (Germany), INFN (Italy), SURF (The Netherlands), PIC (Spain), GridPP (United Kingdom), RRCKI and Yandex LLC (Russia), CSCS (Switzerland), IFIN-HH (Romania), CBPF (Brazil), PL-GRID (Poland) and OSC (USA). We are indebted to the communities behind the multiple open 
source software packages on which we depend.
Individual groups or members have received support from AvH Foundation (Germany),
EPLANET, Marie Sk\l{}odowska-Curie Actions and ERC (European Union), 
Conseil G\'{e}n\'{e}ral de Haute-Savoie, Labex ENIGMASS and OCEVU, 
R\'{e}gion Auvergne (France), RFBR and Yandex LLC (Russia), GVA, XuntaGal and GENCAT (Spain), Herchel Smith Fund, The Royal Society, Royal Commission for the Exhibition of 1851 and the Leverhulme Trust (United Kingdom).



\addcontentsline{toc}{section}{References}
\ifx\mcitethebibliography\mciteundefinedmacro
\PackageError{LHCb.bst}{mciteplus.sty has not been loaded}
{This bibstyle requires the use of the mciteplus package.}\fi
\providecommand{\href}[2]{#2}

\newpage

\newpage
\centerline{\large\bf LHCb collaboration}
\begin{flushleft}
\small
R.~Aaij$^{39}$, 
C.~Abell\'{a}n~Beteta$^{41}$, 
B.~Adeva$^{38}$, 
M.~Adinolfi$^{47}$, 
Z.~Ajaltouni$^{5}$, 
S.~Akar$^{6}$, 
J.~Albrecht$^{10}$, 
F.~Alessio$^{39}$, 
M.~Alexander$^{52}$, 
S.~Ali$^{42}$
% ...

{\footnotesize \it
% ...
$ ^{5}$Clermont Universit\'{e}, Universit\'{e} Blaise Pascal, CNRS/IN2P3, LPC, Clermont-Ferrand, France\\
$ ^{6}$CPPM, Aix-Marseille Universit\'{e}, CNRS/IN2P3, Marseille, France\\
% ...
$ ^{10}$Fakult\"{a}t Physik, Technische Universit\"{a}t Dortmund, Dortmund, Germany\\
% ...
$ ^{38}$Universidad de Santiago de Compostela, Santiago de Compostela, Spain\\
$ ^{39}$European Organization for Nuclear Research (CERN), Geneva, Switzerland\\
% ...
$ ^{41}$Physik-Institut, Universit\"{a}t Z\"{u}rich, Z\"{u}rich, Switzerland\\
$ ^{42}$Nikhef National Institute for Subatomic Physics, Amsterdam, The Netherlands\\
% ...
$ ^{47}$H.H. Wills Physics Laboratory, University of Bristol, Bristol, United Kingdom\\
% ...
$ ^{52}$School of Physics and Astronomy, University of Glasgow, Glasgow, United Kingdom\\
% ...
}
\end{flushleft}

\end{document}